\documentclass[useAMS,fleqn,usenatbib]{mnras}
\usepackage{amssymb} 
\usepackage{amsmath}
\usepackage{graphicx}
\usepackage{color}
\usepackage{hyperref}

\def\jcap{Journal of Cosmology and Astroparticle Physics (JCAP)}

\def\aap{A\&A}
\def\apj{ApJ}
\def\apjs{ApJS}
\def\apjl{ApJL}

\def\mnras{MNRAS}
\def\aj{AJ}

\def\prd{Phys. Rev. D}

\def\aj{AJ}

\def\apj{ApJ}
\def\apjl{ApJ}
\def\apjs{ApJS}

\def\aap{A\&A}

\def\mnras{MNRAS}

\def\prd{Phys.Rev.D}

\def\aj{AJ}

\def\apj{ApJ}
\def\apjl{ApJ}
\def\apjs{ApJS}

\def\aap{A\&A}

\def\mnras{MNRAS}

\def\prd{Phys.Rev.D}

\title[Clustering tomography on the completed BOSS DR12 sample]{The clustering of galaxies in the completed SDSS-III Baryon Oscillation Spectroscopic Survey: Angular clustering tomography and its cosmological implications}
\author[S. Salazar-Albornoz, et al.]{Salvador Salazar-Albornoz$^{1,2}$\thanks{E-mail: ssalazar@mpe.mpg.de}, 
Ariel G. S\'anchez$^2$, Jan Niklas Grieb$^{1,2}$, 
\newauthor Martin Crocce$^{3}$, Roman Scoccimarro$^{4}$, Shadab Alam$^{5,6}$, Florian Beutler$^{7}$,
\newauthor  Joel R. Brownstein$^{8}$, Chia-Hsun Chuang$^{9,10}$, Francisco-Shu Kitaura$^{9}$,
\newauthor  Matthew D. Olmstead$^{11}$, Will J. Percival$^{7}$, Francisco Prada$^{10,12,13}$,
\newauthor Sergio Rodr{\'i}guez-Torres$^{10,14,12}$, Lado Samushia$^{15,16,7}$, Jeremy Tinker$^{4}$, 
\newauthor Daniel Thomas$^7$, Rita Tojeiro$^{17}$, Yuting Wang$^{18,7}$, Gong-bo Zhao$^{18,7}$\\
$^1$Universit\"ats-Sternwarte M\"unchen, Ludwig-Maximilians-Universit\"at M\"unchen, Scheinerstrasse 1, 81679 Munich, Germany\\
$^2$Max-Planck-Institut f\"ur extraterrestrische Physik, Postfach 1312, Giessenbachstr., 85741 Garching, Germany.\\
$^3$Institut de Ci\`encies de l'Espai, IEEC-CSIC, Campus UAB, Carrer de Can Magrans, s/n, E-08193 Bellaterra, Barcelona, Spain.\\
$^4$Center for Cosmology and Particle Physics, Department of Physics, New York University, New York 10003, New York, USA\\
$^5$Department of Physics, Carnegie Mellon University, 500 Forbes Ave., Pittsburgh, PA 15213, USA\\
$^6$The McWilliams Center for Cosmology, Carnegie Mellon University, 500 Forbes Ave., Pittsburgh, PA 15213, USA\\
$^7$Institute of Cosmology \& Gravitation, University of Protsmouth, Dennis Sciama Building, Portsmouth, PO1 3FX, UK\\
$^8$Department of Physics and Astronomy, University of Utah, 115 S 1400 E, Salt Lake City, UT 84112, USA\\
$^9$Leibniz-Institut f{\"u}r Astrophysik (AIP), An der Sternwarte 16, D-14482 Potsdam, Germany\\
$^{10}$Instituto de F\'{\i}sica Te\'orica, (UAM/CSIC), Universidad Aut\'onoma de Madrid,  Cantoblanco, E-28049 Madrid, Spain \\
$^{11}$Department of Chemistry and Physics, King's College, 133 North River St, Wilkes Barre, PA 18711, USA\\
$^{12}$Departamento de F\'{\i}sica Te\'orica, Universidad Aut\'onoma de Madrid, Cantoblanco, 28049, Madrid, Spain \\
$^{13}$Instituto de Astrof\'{\i}sica de Andaluc\'{\i}a (CSIC), Glorieta de la Astronom\'{\i}a, E-18080 Granada, Spain \\
$^{14}$Campus of International Excellence UAM+CSIC, Cantoblanco, E-28049 Madrid, Spain \\
$^{15}$Kansas State University, Manhattan KS 66506, USA\\                        
$^{16}$National Abastumani Astrophysical Observatory, Ilia State University, 2A Kazbegi Ave., GE-1060 Tbilisi, Georgia\\  
$^{17}$School of Physics and Astronomy, University of St Andrews, North Haugh, St Andrews KY16 9SS, UK\\
$^{18}$National Astronomy Observatories, Chinese Academy of Science, Beijing, 100012, P.R. China
}
\begin{document}

\date{Submitted to MNRAS}

\pagerange{\pageref{firstpage}--\pageref{lastpage}} \pubyear{2016}

\maketitle

\label{firstpage}

\begin{abstract}
We investigate the cosmological implications of studying galaxy clustering using a tomographic approach applied to the final BOSS DR12  galaxy sample, including both auto- and cross-correlation functions between redshift shells. We model the signal of the full shape of the angular correlation function, $\omega(\theta)$, in redshift bins using state-of-the-art modelling of non-linearities, bias and redshift-space distortions. We present results on the redshift evolution of the linear bias of BOSS galaxies, which cannot be obtained with traditional methods for galaxy-clustering analysis. We also obtain constraints on cosmological parameters, combining this tomographic analysis with measurements of the cosmic microwave background (CMB) and type Ia supernova (SNIa).  We explore a number of cosmological models, including the standard $\Lambda$CDM model and its most interesting extensions, such as deviations from $w_{\rm{DE}}=-1$, non-minimal neutrino masses, spatial curvature and deviations from general relativity using the growth-index $\gamma$ parametrisation. These results are, in general, comparable to the most precise present-day constraints on cosmological parameters, and show very good agreement with the standard model. In particular, combining CMB, $\omega(\theta)$ and SNIa, we find a value of $w_{\rm{DE}}$ consistent with $-1$ to a precision better than $5\%$ when it is assumed to be constant in time, and better than $6\%$ when we also allow for a spatially-curved Universe.
\end{abstract}


%
\section{Introduction}
Along with measurements of the cosmic microwave background (CMB) and distant type Ia supernovae (SNIa), large galaxy-catalogues tracing the large-scale structure (LSS) of the Universe, have become one of the fundamental observables in observational cosmology.  The most widely used tools for the analysis of the LSS are the so called two-point statistics: the correlation function, and its Fourier counterpart, the power spectrum. These measurements of the clustering of galaxies encode information of both the expansion history of the Universe and the growth of structure. In particular, the baryon acoustic oscillation (BAO) signal imprinted onto these two-point statistics, provides a very robust distance measurement, relative to the sound horizon scale, that can be used to measure the distance-redshift relation probing the expansion history of the Universe.

The BAO signature in the galaxy distribution was simultaneously measured for the first time in 2005 by \cite{Eisenstein:2005aa}, using a spectroscopic subsample of luminous red galaxies (LRGs) of the Sloan Digital Sky Survey \citep[SDSS;][]{York:2000aa}, and by \cite{Cole:2005aa} in the Two-degree Field Galaxy Redshift survey \citep[2dFGRS;][]{Colless:2001aa}. Since then, due to the wealth of information that galaxy surveys provide, much effort has been devoted to design and perform ever larger galaxy-surveys, such as the Baryon Oscillation Spectroscopic Survey \citep[BOSS;][]{Dawson:2013aa}, WiggleZ \citep{Drinkwater:2010aa} and the Dark Energy Survey \citep[DES;][]{The-Dark-Energy-Survey-Collaboration:2005aa}. Supported by this increasing amount of data, substantial work has been devoted to modelling and detecting the BAO signal in two-point statistics and use it for cosmological constraints \citep[e.g.][]{Percival:2007aa,Spergel:2007aa,Reid:2010aa,Blake:2011aa,Sanchez:2014aa,Samushia:2013aa,Anderson:2014aa,Alam:2016aa,Beutler:2016aa}. Future projects, such as the Hobby-Eberly Telescope Dark Energy Experiment \citep[HETDEX; ][]{Hill:2008aa}, the Dark Energy Spectroscopic Instrument \citep[DESI;][]{Levi:2013aa}, the Large Synoptic Survey Telescope \citep[LSST; ][]{LSST-Science-Collaboration:2009aa} and the {\it Euclid} mission \citep{Laureijs:2011aa}, will continue on this path, further improving our understanding of the Universe.

There are two important issues related to the traditional study of LSS that need to be considered. First, in order to use the 3D positions of galaxies, it is necessary to assume a fiducial cosmological model in order to transform the measured angular positions on the sky and redshifts of galaxies into comoving coordinates or distances, a process which could bias the parameter constraints if not treated carefully (see e.g. \citealt{Eisenstein:2005aa} and \citealt{Sanchez:2009aa}). Secondly, in order to obtain a precise measurement of either the correlation function or the power spectrum, usually the whole galaxy sample is used to obtain one measurement, typically averaging over a wide redshift range and assuming that the measurement at the mean redshift is representative of the entire sample, washing out information on the redshift evolution of the structures. 

A simple way to avoid the first issue is to use two-point statistics based only on direct observables, i.e. only angular positions and/or redshifts, such as the angular correlation function $\omega(\theta)$ or the angular power spectrum $C_\ell$. This is done by dividing the sample into redshift bins, or shells, in order to recover information along the line of sight, which otherwise would be lost due to projection effects. Using the clustering in redshift shells solves the second issue of the 3D analysis, providing information on the redshift-evolution of the galaxy-clustering signal, which can be leveraged to put constraints on time-evolving quantities such as the galaxy bias and the growth of structures. Recently, large amount of effort has been committed to develop, test and apply different variation of this methodology \citep[e.g.][]{Crocce:2011aa,Crocce:2011ab,Ross:2011ab,Sanchez:2011aa,de-Simoni:2013aa,Asorey:2012aa,Asorey:2014aa,Di-Dio:2014aa,Salazar-Albornoz:2014aa,Eriksen:2015aa,Eriksen:2015ab,Eriksen:2015ac,Eriksen:2015ad,Carvalho:2016aa}.

This paper extends and applies the {\it clustering tomography} analysis in \cite{Salazar-Albornoz:2014aa} to the final galaxy sample of BOSS. It complements a series of companion papers analysing this sample \citep{Alam:2016aa, Beutler:2017aa, Beutler:2017ab, Chuang:2016aa, Grieb:2017aa, Pellejero-Ibanez:2016aa, Ross:2017aa, Sanchez:2017aa, Sanchez:2017ab,  Satpathy:2016aa, Tinker:2016aa, Vargas-Magana:2016aa, Wang:2016aa, Zhao:2017aa}, and is organised in the following manner: Section \ref{sec:data} outlines our galaxy sample, our measurements and the complementary datasets included in this study. In Section \ref{sec:method} we describe our methodology, including the modelling of the full shape of the angular correlation function in redshift shells, its analytical full covariance matrix, the optimisation of our binning scheme and the performance of this tomographic approach on our set of mock galaxy catalogues. Section \ref{sec:bias} presents our measurements of the redshift evolution of the linear bias of the BOSS galaxy sample, and the impact on cosmological constraints of assuming different models for its evolution. Section \ref{sec:constraints} display our constraint on cosmological parameters for different parameter spaces, obtained combining our measurements of the angular clustering signal in redshift shells with other datasets. Final conclusions are in Section \ref{sec:conclusions}.


\section{The Data}\label{sec:data}
\subsection{The Baryon Oscillation Spectroscopic Survey: DR12}\label{sec:boss}

For our galaxy clustering measurements we use the combined sample of BOSS \citep{Dawson:2013aa} from the final SDSS-III \citep{Eisenstein:2011aa} data release \cite[DR12;][]{Alam:2015aa}, which consists of the combination of the LOWZ and CMASS samples, used separately in previous studies \citep[e.g.][]{Anderson:2014aa,Sanchez:2013aa,Sanchez:2014aa,Beutler:2014ab,Reid:2010aa,Samushia:2014aa,Cuesta:2016aa}, adding up to a sample of over a million galaxies. BOSS galaxies were selected for spectroscopic follow up on the basis of the multi-colour SDSS observations \citep{Gunn:2006aa}, covering the redshift range $0.15<z<0.75$ over an area of $\sim$10000 square degrees. The motivation for the target selection and the algorithms used are described in \cite{Reid:2016aa}. For each target, spectra were obtained using the double-armed BOSS spectrographs \citep{Smee:2013aa}, in order to extract redshifts applying a template-fitting method described in \cite{Bolton:2012aa}.

We used the estimator introduced by \cite{Landy:1993aa} to estimate the angular auto-/cross-correlation function between the redshift shells $p$ and $q$ as
\begin{equation}
	\omega^{(p,q)}(\theta_{i}) = \frac{{\rm{DD}}^{(p,q)}_{i} -{\rm{DR}}^{(p,q)}_{i}-{\rm{DR}}^{(q,p)}_{i}+{\rm{RR}}^{(p,q)}_{i}}{{\rm{RR}}^{(p,q)}_{i}},
\end{equation}
where ${\rm{DD}}_{i}$, ${\rm{DR}}_{i}$ and ${\rm{RR}}_{i}$ are the data-data, data-random and random-random pair counts in the $i$-th $\theta$-bin, respectively. Note that for $p=q$, one obtains the more familiar auto-correlation estimator.

When computing these pair counts, we apply a series of angular weights to account for observational systematic-effects, such as redshift failures, fibre collisions, local stellar density and seeing. These weights are described in detail in \cite{Ross:2017aa}. Each correlation function is measured to a maximum angular separation $\theta_{\rm{max}}(\bar z^{(p,q)})$ corresponding to a physical separation of $\sim180$ $\rm{Mpc}/h$ at the mean redshift of the shell, $\bar z^{(p,q)}$, in the fiducial BOSS DR12 cosmology (see table \ref{tab:concord}) used in analyses based on this galaxy sample \citep{Alam:2016aa, Beutler:2017aa, Beutler:2017ab, Chuang:2016aa, Grieb:2017aa, Pellejero-Ibanez:2016aa, Ross:2017aa, Sanchez:2017aa, Sanchez:2017ab,  Satpathy:2016aa, Vargas-Magana:2017aa, Wang:2016aa, Zhao:2017aa}. We emphasise that the choice of $\theta_{\rm{max}}$ is arbitrary and has no impact on our angular clustering measurements. The total number of bins is chosen to be 18, with varying $\Delta\theta$ corresponding to $\sim9$ $\rm{Mpc}/h$ at the mean redshift of the shell in the fiducial cosmology. These measurements, and their corresponding covariance matrix (see Section \ref{sec:covmod}), have been made publicly available\footnote{\url{https://sdss3.org/science/boss_publications.php}}.

\begin{figure*}
\begin{center}
  \includegraphics[scale=0.4]{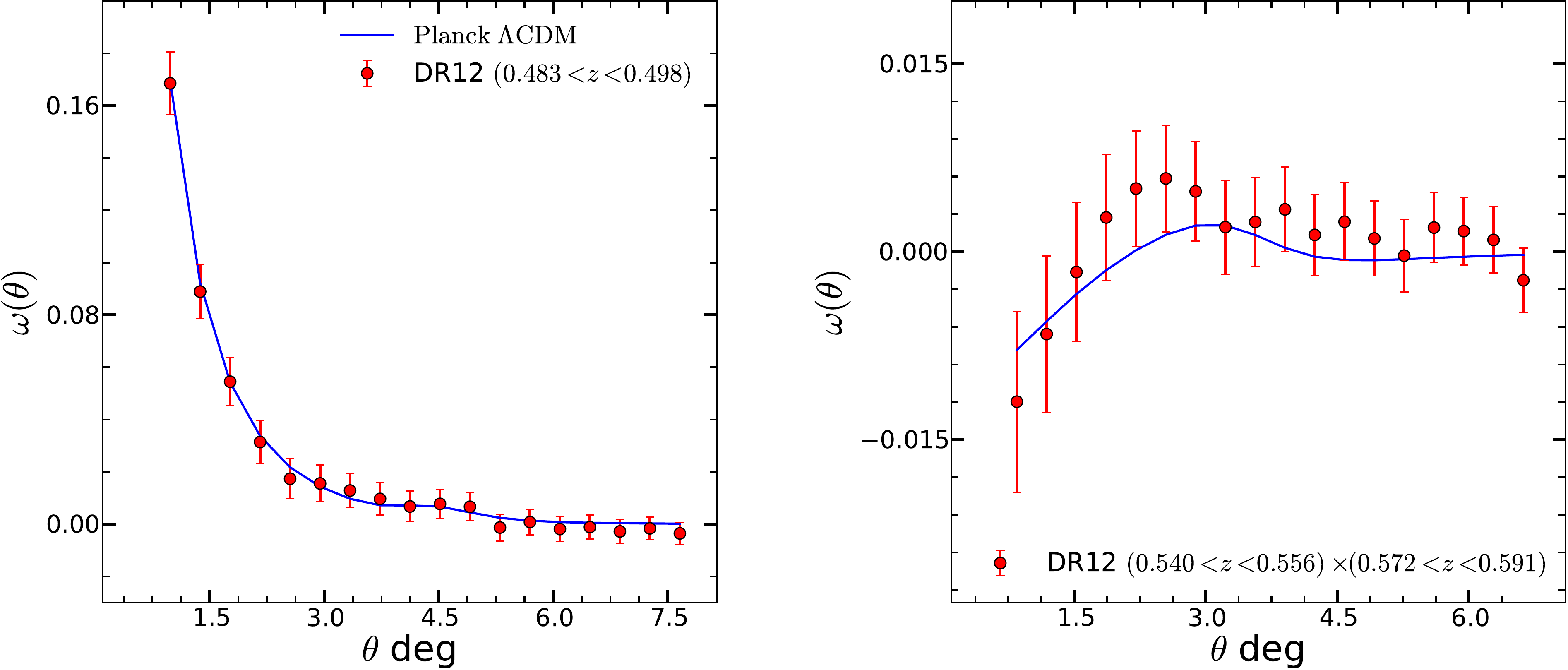}
\end{center}
  \caption{An auto-correlation function (left) and a cross-correlation function (right) between different redshift bins (see key) from the final BOSS galaxy sample. Measurements are shown by red symbols, while the blue line shows the prediction of our model described in Sections \ref{sec:acfmod} and \ref{sec:grptBOSS}, assuming the best-fitting $\Lambda$CDM cosmology from the CMB temperature-anisotropy power spectrum as measured by the {\it Planck} satellite. Errors are derived from our analytical model of the covariance matrix (see Section \ref{sec:covmod}).}
  \label{fig:dr12}
\end{figure*}

For illustration, figure \ref{fig:dr12} shows two measurements on the combined sample (symbols), an auto-correlation function and a cross-correlation function in the left and right panels respectively, for different redshift shells (see key). The blue solid lines correspond to the best-fitting prediction of the model described in Sections \ref{sec:acfmod} and \ref{sec:grptBOSS}, assuming the best-fitting $\Lambda$CDM model from the latest CMB measurements made by the {\it Planck} satellite \citep{Planck-Collaboration:2016ad}.

To test our models for the angular correlation function and its full covariance matrix, we use a set of 1000 MultiDark-Patchy mock catalogues \citep[{\sc md-patchy}; ][]{Kitaura:2016aa}, which are designed to match the characteristics of the final BOSS galaxy sample, following its angular and radial selection function. These mock catalogues also include redshift evolution of galaxy bias and the velocity field (i.e. redshift-space distortions), a crucial characteristic for this analysis. The results of these tests are presented in Section \ref{sec:tests}.

\subsection{Additional data sets}

In order to improve the cosmological constraints obtained in this analysis, in Sections \ref{sec:bias} and \ref{sec:constraints} we combine the information contained in the full shape of $\omega(\theta)$ and its redshift evolution with additional data sets.

We use high-$\ell$ ($\ell=50-2500$) CMB temperature plus the low-$\ell$ ($\ell=2-29$) temperature+polarisation power spectrum, from the latest data release of the {\it Planck} satellite, corresponding to the ``{\it Planck} TT+lowP'' case in \cite{Planck-Collaboration:2016ad}. We refer to this data set simply as ``$\rm{Planck}$'', and to its combination with our $\omega(\theta)$ measurements on BOSS as ``$\rm{Planck}+\omega(\theta)$''.

In addition, we use the luminosity-distance relation information from Type Ia supernova (SNIa). To this end, we use the {\it joint light-curve analysis} compilation \citep[JLA;][]{Betoule:2014aa}, which includes SNIa data from the full SDSS-II \citep{Frieman:2008aa, Kessler:2009aa, Campbell:2013aa} survey and the compilation in \cite{Conley:2011aa}, comprising data from the Supernova Legacy Survey \citep{Astier:2006aa,Sullivan:2011aa}, the {\it Hubble} space telescope \citep{Riess:2007aa,Suzuki:2012aa} and several nearby experiments. We only use this data set in combination with the other two, thus whenever it is included, this is referred to as ``$\rm{Planck}+\omega(\theta)+\rm{SNIa}$''.

\begin{table}
\caption{Cosmological parameters of the BOSS fiducial $\Lambda$CDM cosmology.}
\begin{center}
\begin{tabular}{lcc}
\hline\hline
\mbox{Cosmological constant density parameter} & $\Omega_\Lambda$ & $0.69$\\
\mbox{Matter density parameter} &  $\Omega_{\rm m}$ & $0.31$\\
\mbox{Baryonic density parameter} & $\Omega_{\rm b}$ & $0.048$\\
\mbox{Dark energy equation of state} & $w_{DE}$ & $-1.0$\\
\mbox{Hubble constant $\left(\mbox{km s}^{-1}\right.$Mpc$\left.^{-1}\right)$} & $H_0$ & $67.6$\\
\hline\hline
\end{tabular}
\end{center}
\label{tab:concord}
\end{table}

\section{Methodology}\label{sec:method}

We base our methodology on that in \citet{Salazar-Albornoz:2014aa}, extending for the inclusion of cross-correlations between different shells. This description of $\omega(\theta)$ includes local and non-local bias effects, non-linear growth of structures and redshift-space distortions, but neglects relativistic effects such as the integrated Sachs-Wolfe effect, lensing, and magnification bias \citep{Yoo:2009aa,Yoo:2009ab,Bonvin:2011aa,Challinor:2011aa,Cardona:2016aa}, whose effect on the clustering measurements from BOSS should be negligible.
In Section \ref{sec:acfmod} we model the projection of the clustering signal onto angular coordinates for a general model. In Section \ref{sec:grptBOSS} we show the particular model for the 3D clustering of galaxies used in this analysis. After that, our analytical model for the full covariance matrix of $\omega(\theta)$ is shown in Section \ref{sec:covmod}. Using these tools we optimise the binning scheme applied to BOSS in Section \ref{sec:opti}, to finally test this methodology in Section \ref{sec:tests}.

\subsection{Modelling $\bomega(\btheta)$}\label{sec:acfmod}

Given the redshift shells $p$ and $q$, the angular auto-/cross-correlation function is given by,
\begin{equation}
  \omega^{(p,q)}(\theta) = \int dz_1\phi^{p}(z_1) \int dz_2\phi^{q}(z_2) \xi(s,\mu_s), 
  \label{eq:acfmodel}
\end{equation}
where $\phi^{p}(z)$ and $\phi^{q}(z)$ are the normalised selection functions of the shells $p$ and $q$ respectively, and $\xi(s,\mu_s)$ is the anisotropic spatial correlation function at the mean redshift $\bar z^{(p,q)}$.  We also need expressions for the comoving separation $s$ and $\mu_s\equiv\cos\varphi$, the cosine of the angle $\varphi$ between the separation vector and the line of sight, as a function of $\lbrace z_1, z_2,\theta\rbrace$. 

Assuming that the geometry of the Universe is described by the FRW metric, the line-of-sight comoving distance to a given redshift $z$ is given by,
\begin{equation}
   D_{\rm{C}} (z)= D_{\rm{H}} \chi(z),
   \label{eq:Dc}
\end{equation}
where $D_{\rm{H}}\equiv\frac{c}{H_0}$ is the Hubble distance, $H_0$ is the value of the Hubble constant today, and $\chi(z)$ is given by
\begin{equation}
   \chi(z)= \int_0^z \frac{dz^\prime}{E(z^\prime)},
   \label{eq:chi}
\end{equation}
defining $E(z)\equiv \frac{H(z)}{H_0}$. On the other hand, the transverse comoving distance, defined as the comoving distance we would infer between two objects at the same redshift knowing their angular and comoving separation, is given by,
 \begin{equation}
    D_{{\rm{M}}}(z) = \left\lbrace \begin{array}{cc}
   					\frac{D_{\rm{H}}}{\sqrt{|\Omega_{\rm{K}}|}}S_{\rm{K}}\left[\chi(z)\right] & \Omega_{\rm{K}} \neq 0\\
					&\\
					D_{\rm{H}}S_{\rm{K}}\left[\chi(z)\right] & \Omega_{\rm{K}} = 0
   				\end{array} \right.,
    \label{eq:Dm}
 \end{equation}
 where $\Omega_{\rm{K}}$ is the curvature density parameter today, and $S_{\rm{K}}\left[\chi(z)\right]$ is defined as
\begin{equation}
   S_{\rm{K}}\left[\chi(z)\right] = \left\lbrace \begin{array}{cc} 
				\sinh\left(\sqrt{\Omega_{\rm{K}}}\chi(z)\right) & \Omega_{\rm{K}} > 0\\ 
                   		&\\
                   		\chi(z) & \Omega_{\rm{K}} = 0\\
                   		&\\
                   		\sin\left(\sqrt{|\Omega_{\rm{K}}|}\chi(z)\right) & \Omega_{\rm{K}} < 0\\ 
    			     \end{array} \right. . 
    \label{eq:Sk}
\end{equation}

With this, the comoving separation between two objects (galaxies), observed by us at different redshifts, and with an angular separation $\theta$ on the sky, $s(z_1,z_2,\theta)$, is given by
\begin{equation}
   s(z_1,z_2,\theta) = \left\lbrace \begin{array}{cc}
   					\frac{D_{\rm{H}}}{\sqrt{|\Omega_{\rm{K}}|}}S_{\rm{K}}\left[\chi_{(1,2)}\right] & \Omega_{\rm{K}} \neq 0\\
					&\\
					D_{\rm{H}}S_{\rm{K}}\left[\chi_{(1,2)}\right] & \Omega_{\rm{K}} = 0
   				\end{array} \right.,
   \label{eq:s}
\end{equation}
where $\chi_{(1,2)}$ is given by equation (\ref{eq:chi}) as if object 1 were observing object 2 at the time the light observed by us was emmited, and $S_{\rm{K}}\left[\chi_{(1,2)}\right]$ can be obtained from the spherical cosine rule (generalised for positive and negative curvature) as \citep{Peacock:1999aa,Liske:2000aa}, 
\begin{equation}
  \begin{split}
    S_{\rm{K}}^2\left[\chi_{(1,2)}\right] = S_{\rm{K}}^2\left[\chi(z_1)\right]C_{\rm{K}}^2\left[\chi(z_2)\right] +  S_{\rm{K}}^2\left[\chi(z_2)\right]C_{\rm{K}}^2\left[\chi(z_1)\right]\\ 
    - {\rm sgn}(\Omega_{\rm{K}})S_{\rm{K}}^2\left[\chi(z_1)\right]S_{\rm{K}}^2\left[\chi(z_2)\right]\sin^2\theta \\
    - 2 S_{\rm{K}}\left[\chi(z_1)\right]S_{\rm{K}}\left[\chi(z_2)\right]C_{\rm{K}}\left[\chi(z_1)\right]C_{\rm{K}}\left[\chi(z_2)\right]\cos\theta, 
  \end{split}
  \label{eq:Sks}
\end{equation}
where $C_{\rm{K}}$ is defined as
\begin{equation}
   C_{\rm{K}}\left[\chi(z)\right] = \left\lbrace \begin{array}{cc} 
				\cosh\left(\sqrt{\Omega_{\rm{K}}}\chi(z)\right) & \Omega_{\rm{K}} > 0\\ 
                   		&\\
                   		1 & \Omega_{\rm{K}} = 0\\
                   		&\\
                   		\cos\left(\sqrt{|\Omega_{\rm{K}}|}\chi(z)\right) & \Omega_{\rm{K}} < 0\\ 
    			     \end{array} \right. .  
    \label{eq:Ck}
\end{equation}
Note that when $\Omega_{\rm{K}}=0$, equation (\ref{eq:s}) reduces to the well known Euclidean expression,
\begin{equation}
  s(z_1,z_2,\theta) = \sqrt{D_{\rm{C}}^2(z_1)+D_{\rm{C}}^2(z_2)-2D_{\rm{C}}(z_1)D_{\rm{C}}(z_2)\cos\theta}.
  \label{eq:s_flat}
\end{equation}
The difference in using equation (\ref{eq:s}), compared to equation (\ref{eq:s_flat}) with the correct form of $D_{{\rm{M}}}$, is of the order of few per-cent when $\Omega_{\rm{K}} \in [-0.2,0.2]$. This difference translates directly into a shift of the same order on the estimation of the BAO position, which can be significant for a sample able to achieve percent-level precision.

Similarly, using the (generalised) spherical sine rule, we can find a simple expression for $\sin\varphi$, the sine of the angle between the separation vector and the line of sight, which is given by
\begin{equation}
  \sin\varphi = \frac{S_{\rm{K}}\left[\chi(z_1)\right]S_{\rm{K}}\left[\chi(z_2)\right]\sin\theta}{S_{\rm{K}}\left[\chi_{(1,2)}\right]S_{\rm{K}}\left[\chi^\prime\right]},
  \label{eq:sinvarphi}
\end{equation}
where $D_{\rm{H}}\chi^\prime$ is the line-of-sight comoving distance between the observer and the mid-point of the separation vector. Now, we only need $S_{\rm{K}}\left[\chi^\prime\right]$ to calculate $\sin\varphi$, and then take\footnote{Note that we can drop the $\pm$, and take the positive solution, since redshift-space distortions are symmetric around the line-of-sight.} $\mu_s = \sqrt{1-\sin^2\varphi}$. Since $S_{\rm{K}}\left[\chi^\prime\right]$ is the median of the spherical triangle defined by $z_1$, $z_2$, $\theta$ and the observer, using Stewart's theorem we have the relation
\begin{equation}
     C_{\rm{K}}\left[\chi^\prime\right] = \frac{C_{\rm{K}}\left[\chi(z_1)\right]+ C_{\rm{K}}\left[\chi(z_2)\right]}{2C_{\rm{K}}\left[\frac{\chi_{(1,2)}}{2}\right]}.
  \label{eq:stweart}
\end{equation}
Note that this relation only works for $\Omega_{\rm{K}}\neq0$, and gives a trivial solution for a flat geometry. In the case when $\Omega_{\rm{K}}=0$, we should use 
\begin{equation}
   \mu_s  = \frac{D^2_{{\rm{M}}}(z_2)-D^2_{{\rm{M}}}(z_1)}{s \ \sqrt{D^2_{{\rm{M}}}(z_1) + D^2_{{\rm{M}}}(z_2) + 2D_{{\rm{M}}}(z_1)D_{{\rm{M}}}(z_2)\cos\theta}}.
   \label{eq:flat_mu}
\end{equation}
The difference between deriving $\mu_s$ using (\ref{eq:sinvarphi}) for  $\Omega_{\rm{K}}\neq0$, compared to using equation (\ref{eq:flat_mu}) with the correct form of $D_{{\rm{M}}}$ for any value of $\Omega_{\rm{K}}$, is less than $0.2\%$ for the range of angular and redshift separations we are considering, while the second case is significantly faster to compute. For this reason, we compute $\mu_s$ using equation (\ref{eq:flat_mu}) in our analysis later on. 

When comparing the model for $\omega(\theta)$ with measurements, it is important to take into account the effect of the binning in $\theta$. Measurements are not done over a single angle $\theta$, but correspond to the average over a bin centred on $\theta$ with a bin-width $\Delta\theta$. In order to avoid systematic effects such as a shift in the BAO peak determination, we consider in our analysis the bin-averaged angular correlation function, evaluated at the bin $\theta_i$, given by

\begin{equation}
 \omega(\theta_i) = \frac{1}{\Delta\Omega_i} \int_{\Delta\Omega_i}d\Omega \ \omega(\theta), \label{eq:binwt}
\end{equation}
where $\Delta\Omega_i$ is the solid angle given by
\begin{equation}
 \Delta\Omega_i = 2\pi\int_{\theta_i-\Delta\theta/2}^{\theta_i+\Delta\theta/2}d\theta^\prime \sin\theta^\prime. \label{eq:DOmega}
\end{equation}

\subsection{Anisotropic galaxy clustering}\label{sec:grptBOSS}

For the anisotropic spatial correlation function $\xi(s,\mu_s)$, we use the same framework as in \cite{Sanchez:2017aa} (see also \citealt{Grieb:2017aa} for Fourier space), which is inspired in gRPT \citep{Crocce:2016aa} for the clustering of matter in real space, and describes galaxy bias and redshift-space distortions (RSD) with four parameters: two local bias parameters $b_1$ and $b_2$, a non-local bias parameter $\gamma^-_3$, and one parameter for the {\it fingers-of-god} effect, $a_\rmn{vir}$, characterising the kurtosis of the velocity distribution within virialised structures. In order to correctly model $\omega^{(p,q)}(\theta)$ we need to compute the line-of-sight projection of $\xi(s,\mu_s;\bar z^{(p,q)})$ as in equation (\ref{eq:acfmodel}).  For this, we need to consider that the galaxy bias evolves with redshift, as well as the signal of the RSD and the non-linear growth of structures. In practice, this means that the nuisance parameters of our model, $\lbrace b_1, b_2, \gamma_3^-, a_{\rm{vir}}\rbrace$, will have different values at different redshifts. Here we describe how we treat the redshift evolution of $\xi(s,\mu_s)$, and refer the reader to the papers mentioned above for a more detailed description of the model in configuration and Fourier space.

We assume that the redshift evolution of $\xi(s,\mu_s;\bar z^{(p,q)})$, including all effects considered here, can be safely neglected within a single measurement due to their smooth and monotonic evolution with $z$ (see Sec. 3 in \citealt{Salazar-Albornoz:2014aa} and references therein). This means quantities evaluated at $\bar z^{(p,q)}$ are effectively a combination of their mean values within the boundaries of the redshift-shells $p$ and $q$, weighed by the corresponding $\phi^p(z)$ and $\phi^q(z)$.

For the linear galaxy bias parameter $b_1$, we test three well motivated models. First, the vast majority of galaxies in BOSS are old passively-evolving galaxies \citep{Leauthaud:2016aa}, this motivates the use of the model in \cite{Fry:1996aa} (hereafter F96), given by
\begin{equation}
     b_1\left(\bar z^{(p,q)}\right) = 1 + (b_1-1)\frac{D(z_{\rm{ref}})}{D\left(\displaystyle\bar z^{(p,q)}\right)},
  \label{eq:biasFry}
\end{equation}
where $D(z)$ is the linear-theory growth factor. On the other hand, it has been shown empirically that the clustering amplitude of CMASS galaxies does not evolve significantly with redshift \citep{Reid:2014aa, Saito:2016aa}. If the amplitude of the matter density fluctuations evolves (in the linear regime) with the linear growth factor, then the galaxy bias needs to evolve as
\begin{equation}
     b_1\left(\bar z^{(p,q)}\right) = b_1\frac{D(z_{\rm{ref}})}{D\displaystyle\left(\bar z^{(p,q)}\right)},
  \label{eq:biasCGC}
\end{equation}
in order to keep the amplitude of the galaxy-clustering signal constant. This model is referred to as the constant galaxy-clustering model (hereafter CGC). These two models relate the evolution of the galaxy bias with the linear growth factor, which could lead to biases in the cosmological parameters if the models are not correct. For this reason, we also test a simple linear model that does not depend on the cosmology, given by
\begin{equation}
     b_1\left(\bar z^{(p,q)}\right) = b_1 + b^\prime\left(\bar z^{(p,q)} - z_{\rm{ref}}\right),
  \label{eq:linb}
\end{equation}
where $b^\prime$ is an extra nuisance parameter to be fit when using this model. We do not expect a redshift dependence of the quadratic bias parameter $b_2$.

The redshift evolution for the non-local bias parameter is given by
\begin{equation}
     \gamma_3^-\left(\bar z^{(p,q)}\right) = \gamma_3^-\frac{D(z_{\rm{ref}})}{D\displaystyle\left(\bar z^{(p,q)}\right)},
  \label{eq:gamma3}
\end{equation}
while $a_{\rm{vir}}$ evolves with redshift as 
\begin{equation}
     a_{\rm{vir}}\left(\bar z^{(p,q)}\right) = a_{\rm{vir}}\left(\frac{D\left(\bar z^{(p,q)}\right)}{D(z_{\rm{ref}})}\right)^2.
  \label{eq:afog}
\end{equation}

\begin{figure}
\begin{center}
     \includegraphics[scale=0.6]{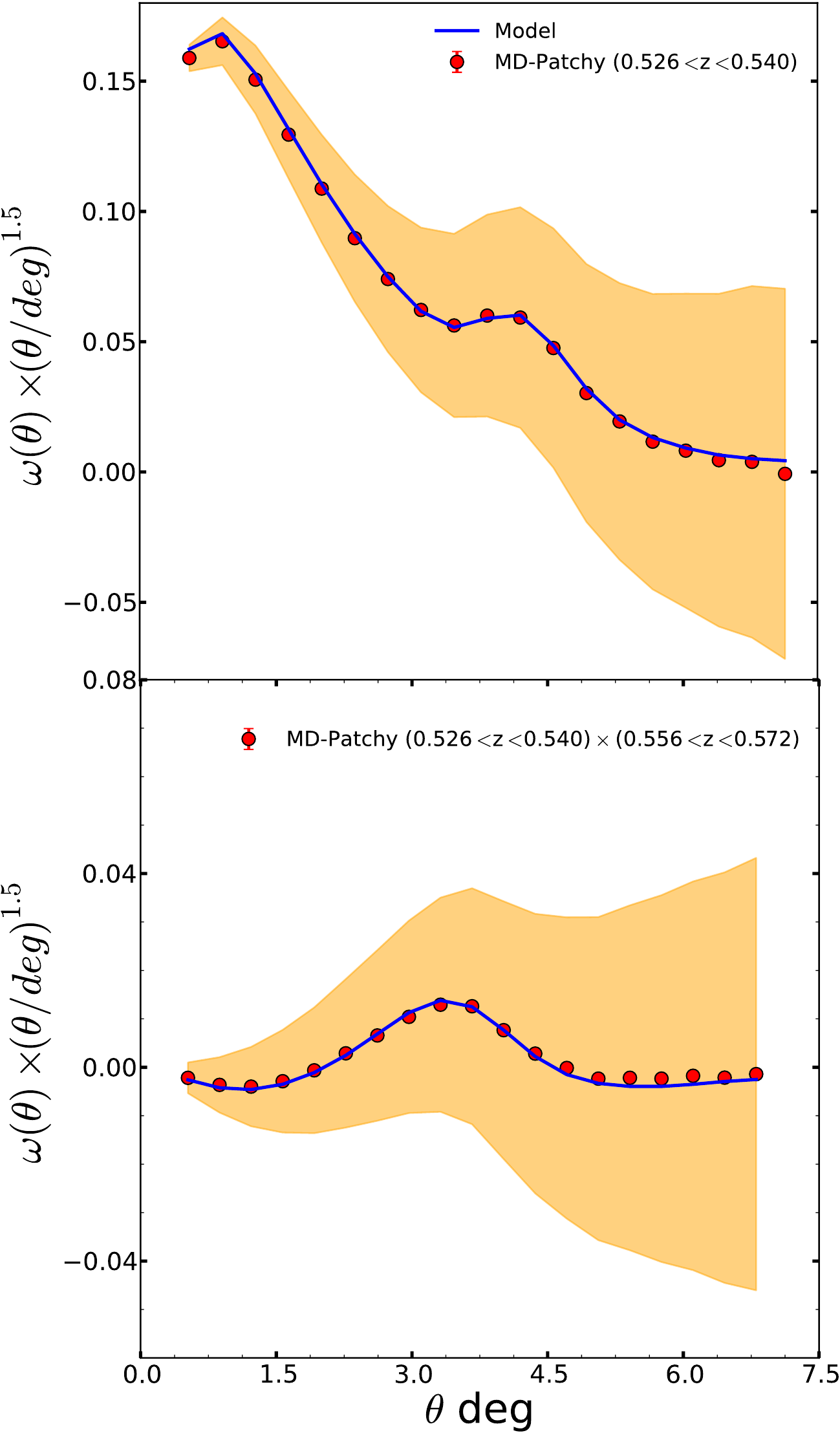}
 \end{center}
   \caption{Comparison between the best-fitting model (blue solid line) and the mean of the 1000 {\sc md-patchy} (symbols). The top panel shows an auto-correlation function, and the bottom panel a cross-correlation function. In both panels the colour band shows the dispersion corresponding to a single realisation.}
  \label{fig:model_patchy}
\end{figure}

Figure \ref{fig:model_patchy} shows a comparison between the best-fitting model (blue solid line) and the mean of the 1000 {\sc md-patchy} (symbols). Here we use the bias model in eq. (\ref{eq:biasFry}), and the true underlying linear matter power spectrum $P(k)$. The upper panel shows one of the auto-correlation functions measured, and the lower panel a cross-correlation function. In both panels the colour band shows the dispersion corresponding to a single realisation. 

\subsection{Analytical model for the full covariance matrix}\label{sec:covmod}

Noise in covariance matrix estimates from mock catalogs propagates to the recovered likelihood of  cosmological parameters, leading to an increase in the final errors in those parameters \citep{Dodelson:2013aa,Taylor:2013aa,Taylor:2014aa,Percival:2014aa}. These uncertainties, and so their correction, depend on the number of mock catalogs used to estimate the covariance matrix, the number of bins in the data vector, and the number of parameters to be constrained using this matrix. In order to keep this extra source of uncertainty below $1\%$, it would be necessary to measure $\sim10^5$ independent mock catalogs. Therefore, we use an analytical form instead, that has been shown to be in excellent agreement with N-body simulations \citep{Crocce:2011aa,Salazar-Albornoz:2014aa}.

The full bin-averaged covariance matrix can be obtained as
\begin{equation}
   \begin{split}
     	{\rm{Cov}}_{i,j}^{(m,n),(p,q)} = \sum_{\ell,\ell^\prime \geq 2}\left( \frac{2\ell+1}{4\pi} \right)^2\left[ \hat L_\ell\left(\cos\theta_{i}\right) \right. \\
         \hat L_{\ell^\prime}\left(\cos\theta_{j}\right)\left.{\rm{Cov}}_{\ell,\ell^\prime}^{(m,n),(p,q)} \right],
     \end{split}
   \label{eq:covtheta} 
\end{equation}
where $\lbrace m,n,p,q\rbrace$ denote for every redshift shell in our configuration, $\hat L_\ell\left(\cos\theta_{i}\right)$ is the bin-averaged Legendre polynomial of $\ell$-th order in the solid angle $\Delta\Omega_{i}$ defined by the angular bin $\theta_{i}$ as
\begin{equation}
   \begin{split}
      \hat L_\ell(\cos\theta_\rmn{i}) & =  \frac{1}{\Delta\Omega_{\rmn i}} \int_{\Delta\Omega_{\rmn i}} d\Omega L_\ell(\cos\theta_\rmn{i}) \\
                                 & =  \frac{2\pi}{\Delta\Omega_i}\frac{1}{2\ell+1}\left[ L_{\ell-1}\left(\cos(\theta_\rmn{i}+\Delta\theta/2)\right) \right. \\
                                 & \hskip 0.4cm - L_{\ell+1}\left(\cos(\theta_\rmn{i}+\Delta\theta/2)\right) - L_{\ell-1}\left(\cos(\theta_\rmn{i}-\Delta\theta/2)\right)\\
                                 & \hskip 0.4cm    \left. + L_{\ell+1}\left(\cos(\theta_\rmn{i}-\Delta\theta/2)\right) \right],
  \end{split}
   \label{eq:Legend}
\end{equation}
 and ${\rm{Cov}}_{\ell,\ell^\prime}^{(m,n),(p,q)}$ is the covariance matrix of the angular power spectrum $C_\ell$ which, assuming that the density field is a Gaussian random field, is given by
\begin{equation}
  {\rm{Cov}}_{\ell,\ell^\prime}^{(m,n),(p,q)} = \delta_{\ell\ell^\prime}\frac{\hat C_\ell^{(m,p)}\hat C_\ell^{(n,q)} + \hat C_\ell^{(m,q)}\hat C_\ell^{(n,p)}}{f_{\rm{sky}}(2\ell+1)}.
  \label{eq:covcl}
\end{equation}
Here, $\delta_{xy}$ is the kronecker delta function, and $\hat C_\ell$ is the angular galaxy-power-spectrum, as it would be observed
\begin{equation}
  \hat C_\ell^{(p,q)} = C_\ell^{(p,q)} + \frac{\delta_{pq}}{\bar n^p},
  \label{eq:obscl}
\end{equation}
where $\bar n^{p}$ is the mean number of galaxies per steradian in the redshift shell $p$, and $1/\bar n^p$ is the shot-noise contribution to auto-correlations.

Assuming the BOSS fiducial cosmology, we compute the redshift-space galaxy $C_\ell^{(p,q)}$ using the {\sc class} code \citep{Blas:2011aa}, taking into account the specific radial selection, and a linear bias evolution that fits that of the data (see Section \ref{sec:bias}) normalised to the corresponding $\sigma_8$ in this cosmology. 

\begin{figure}
 \includegraphics[scale=0.45]{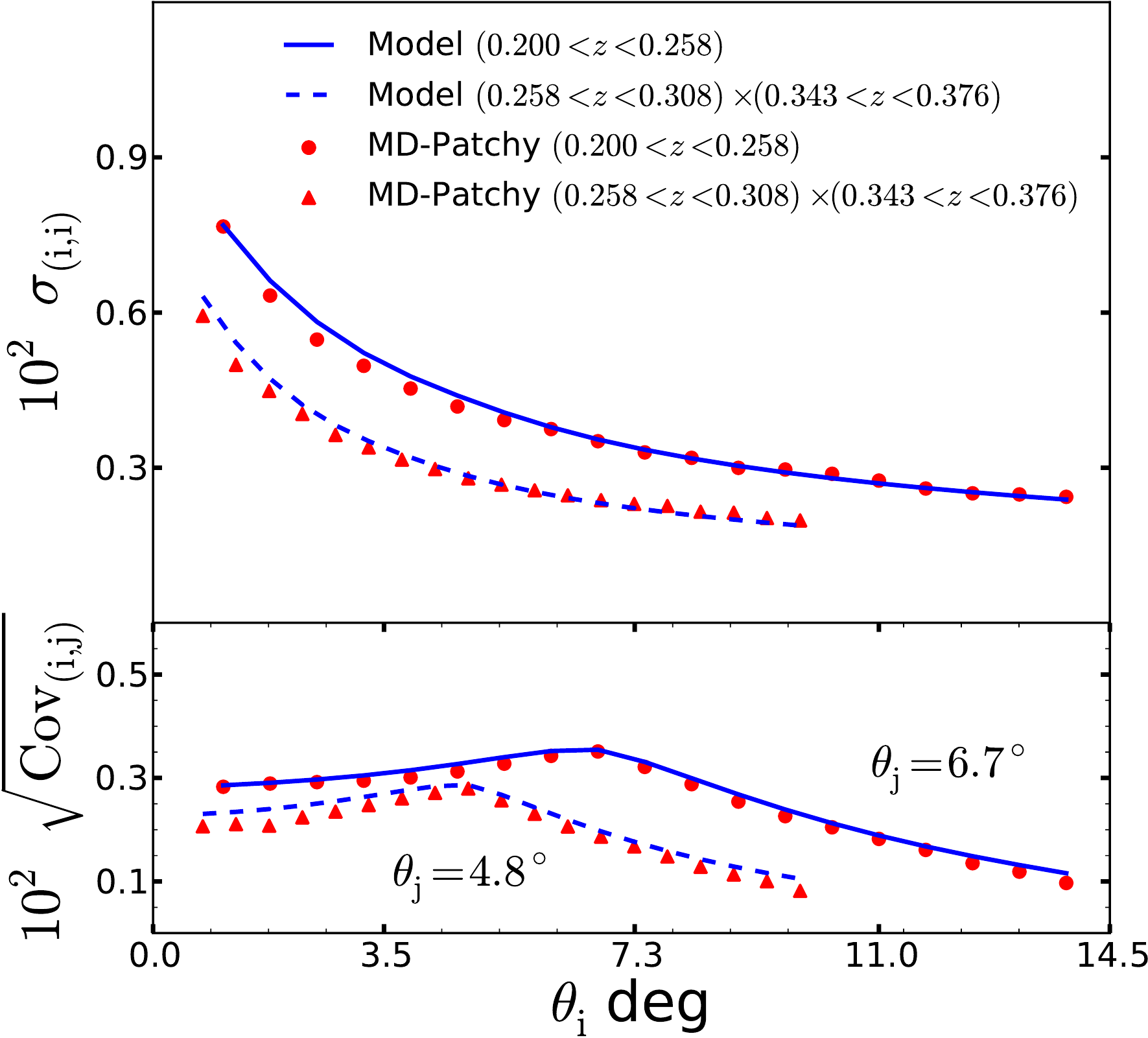}
 \caption{Comparison between sections of the model (dashed and solid lines) and the estimate from the mock catalogues (symbols). The upper panel shows the square root of the diagonal of two sub-matrices corresponding to an auto-correlation and a cross-correlation function measurement (see key). The bottom panel shows the square root of vertical cuts of the same sub-matrices at a fixed $\theta_{j}$ bin.}
 \label{fig:cov}
\end{figure}

For consistency, since we do not know a priori the true cosmology of the Universe, we use this covariance matrix for the data analysis and all the tests performed on our mock catalogues, irrespective of their true fiducial cosmology. For illustration, Figure \ref{fig:cov} shows a comparison of some sections of the covariance-matrix model (dashed and solid lines) against one estimated from the mocks (symbols). The upper panel shows the square root of the diagonal of two sub-matrices corresponding to an auto-correlation and a cross-correlation function measurement (see key), and the bottom panel shows the square root of vertical cuts of the same sub-matrices at a fixed $\theta_{j}$ bin.

\subsection{Redshift binning optimisation}\label{sec:opti}

The binning scheme in redshift shells is a significant variable to consider for our analysis. Thinner shells result in  a sharper BAO feature, at the expense of increasing the statistical uncertainties (due to the smaller number of objects) and the correlation between different shells. Thicker shells, on the other hand, improve the statistical errors, while lowering the BAO signal because it is projected over a wider range of angular scales.

\begin{figure}
 \includegraphics[scale=0.45]{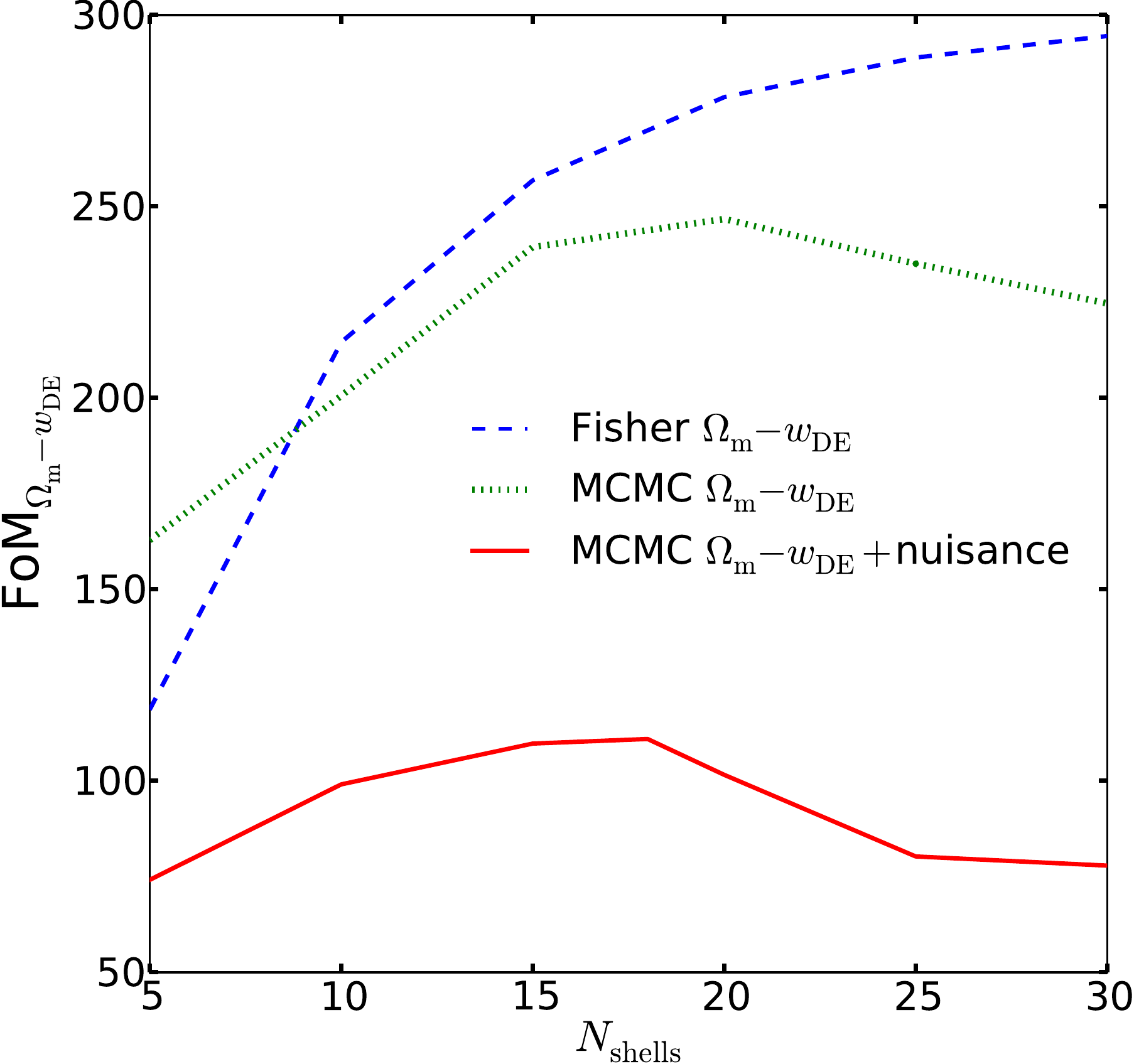}
 \caption{Figure-of-Merit constraining $\Omega_{{{\rm m}}} - w_{{\rm{DE}}}$ as a function of the number of shells for the combined BOSS sample. The blue dashed line shows the prediction using the Fisher matrix-information technique, the green dashed line shows the prediction from the MCMC analysis when only the cosmological parameters are allowed to vary, and the red solid line shows that of the case where we also include the model nuisance-parameters in the MCMC analysis.}
 \label{fig:FoM}
\end{figure}

To maximise the constraining power of our analysis, we optimise the number and the width of the redshift shells we use. Our optimisation is based on the binning strategy in \cite{Di-Dio:2014aa}, which defines the width $\Delta z$ of each shell in such a way that all of them have the same number of galaxies. This results in a constant shot-noise in all our measurements, which is the main contributor to the covariance matrix in a sample with the number density of BOSS. In this procedure we use a smoothed version of the radial number counts, $N(z)$, in order to avoid our binning to be affected by the clustering itself. 

The criteria to define the optimal binning scheme is to maximise the Figure-of-Merit (FoM) in the $\Omega_{{\rm m}}-w_{\rm{DE}}$ plane, defined as
\begin{equation}
   {\rm{FoM}}_{w_{\rm{DE}},\Omega_{{\rm m}}} = \frac{1}{\sqrt{{\rm{det}}[{\rm{Cov}}(w_{\rm{DE}},\Omega_{{\rm m}})]}},
   \label{eq:optiFoM}
\end{equation}
where ${\rm{det}}[{\rm{Cov}}(w_{\rm{DE}},\Omega_{{\rm m}})]$ is the determinant of the covariance matrix between the two parameters being constrained. We only use the cosmological information encoded in the full shape of $\omega(\theta)$ for this purpose. 

First, we test our optimisation procedure using only auto-correlations, exploring two different methods to compute the FoM for a given configuration:
\begin{itemize}
   \item[(i)] a Fisher information-matrix analysis,
   \item[(ii)] a Markov chain Monte Carlo (MCMC) analysis, based on \cite{Salazar-Albornoz:2014aa}, using synthetic data. 
\end{itemize} 
Both methods are performed using our model of the full covariance matrix of $\omega(\theta)$, and taking into account the specific characteristics of BOSS (i.e. angular and radial selection function). Thus, the optimal binning scheme found here is specific for BOSS, and does not apply to other galaxy surveys. We perform two versions of the MCMC analysis: one varying only the cosmological parameters, and another one where we also include the nuisance parameters of our model. 

Figure \ref{fig:FoM} shows the obtained values of the FoM for these three tests, as a function of the number of redshift shells, $N_{\rm{shells}}$. The blue dashed line corresponds to the predictions from the Fisher matrix analysis, the green dashed line shows the predictions from the MCMC analysis when only $w_{\rm{DE}}$ and $\Omega_{\rm m}$ are allowed to vary, and the red solid line shows the results of the case where we also include the model nuisance-parameters in the MCMC analysis. While the Fisher analysis always predicts a monotonically higher FoM as the number of shells increases, none of the MCMC analyses shows this behaviour, where the value of the FoM has a maximum and then decays. This might be explained by the fact that the Fisher matrix analysis approximates the shape of the posterior distribution by a multivariate Gaussian, which in reality is not correct for this combination of parameters. Thus, as $N_{\rm{shells}}$ increases, the reduction of the posterior-distribution surface (which is what the FoM is actually estimating) is not equal for both methods. This, in the Fisher analysis case, could compensate the lost of information in the regime where the shot noise dominates (high $N_{\rm{shells}}$).

Regarding the two different MCMC analysis, it is clear that the inclusion of the nuisance parameters also changes the optimal value of $N_{\rm{shells}}$. For this reason, in the following we only use the ``$w_{\rm{DE}}-\Omega_{\rm m}+$nuisance'' method.

Next, we extend the analysis of the optimal binning-scheme by including the cross-correlations between different redshift shells, imposing two conditions: 
\begin{itemize}
	\item[(i)] as before, each redshift shell must contain the same number of galaxies and, 
	\item[(ii)] for each redshift shell, we include as many cross-correlations with subsequent redshift-shells as necessary to reach at least $120{\rm{Mpc}}/h$ (in the BOSS fiducial cosmology), i.e. past the BAO scale in the line-of-sight direction. The cross-correlation signal is already very close to zero, thus including measurements of redshift shells that are further apart than the \textit{zero-crossing} point of $\xi$ does not add extra information in our case.
\end{itemize}
In this test we also find that the maximum is consistent with the previous tests, but the value of the FoM increases by a factor $\sim2$, with respect to the case where we only use auto-correlations.

As a result, the optimal binning scheme for the combined sample of BOSS is set to $18$ redshift-shells, each of them with $\sim70000$ objects. The redshift limits of the optimal binning for the combined sample are listed in table \ref{tab:zlim}. In Section \ref{sec:bias} we show that, in order to obtain robust cosmological constraints, we need to exclude the last three redshift shells at $z\gtrsim0.6$. For this reason, the final configuration consists of $40$ measurements in total, $15$ auto-correlation functions and $25$ cross-correlation functions, as shown in Figure \ref{fig:confmat} in matrix form.

\subsection{Model performance on mock catalogues}\label{sec:tests}

\begin{figure*}
\begin{center}
   $\begin{array}{ccc}
      \includegraphics[scale=0.305]{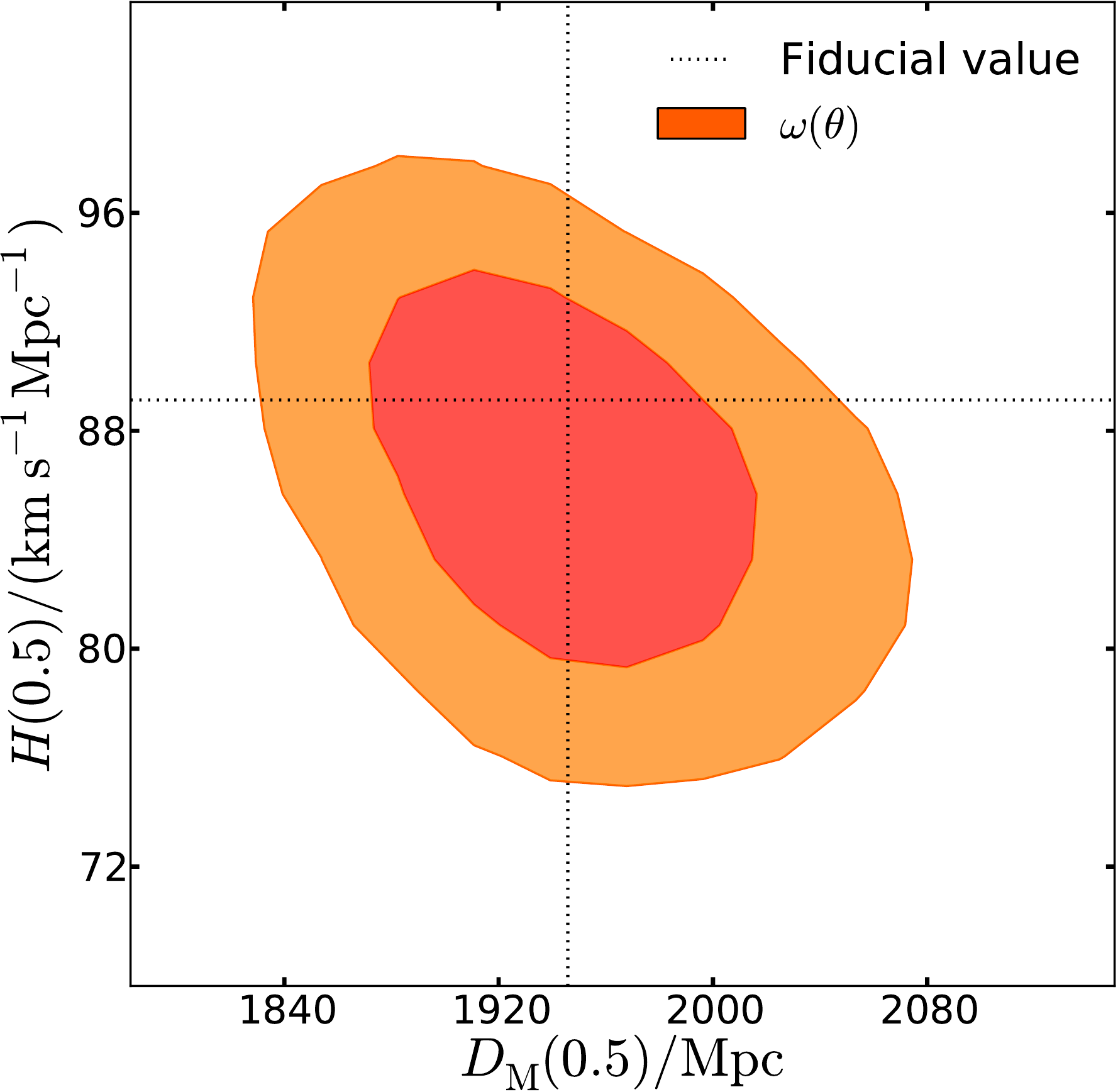} &
      \includegraphics[scale=0.311]{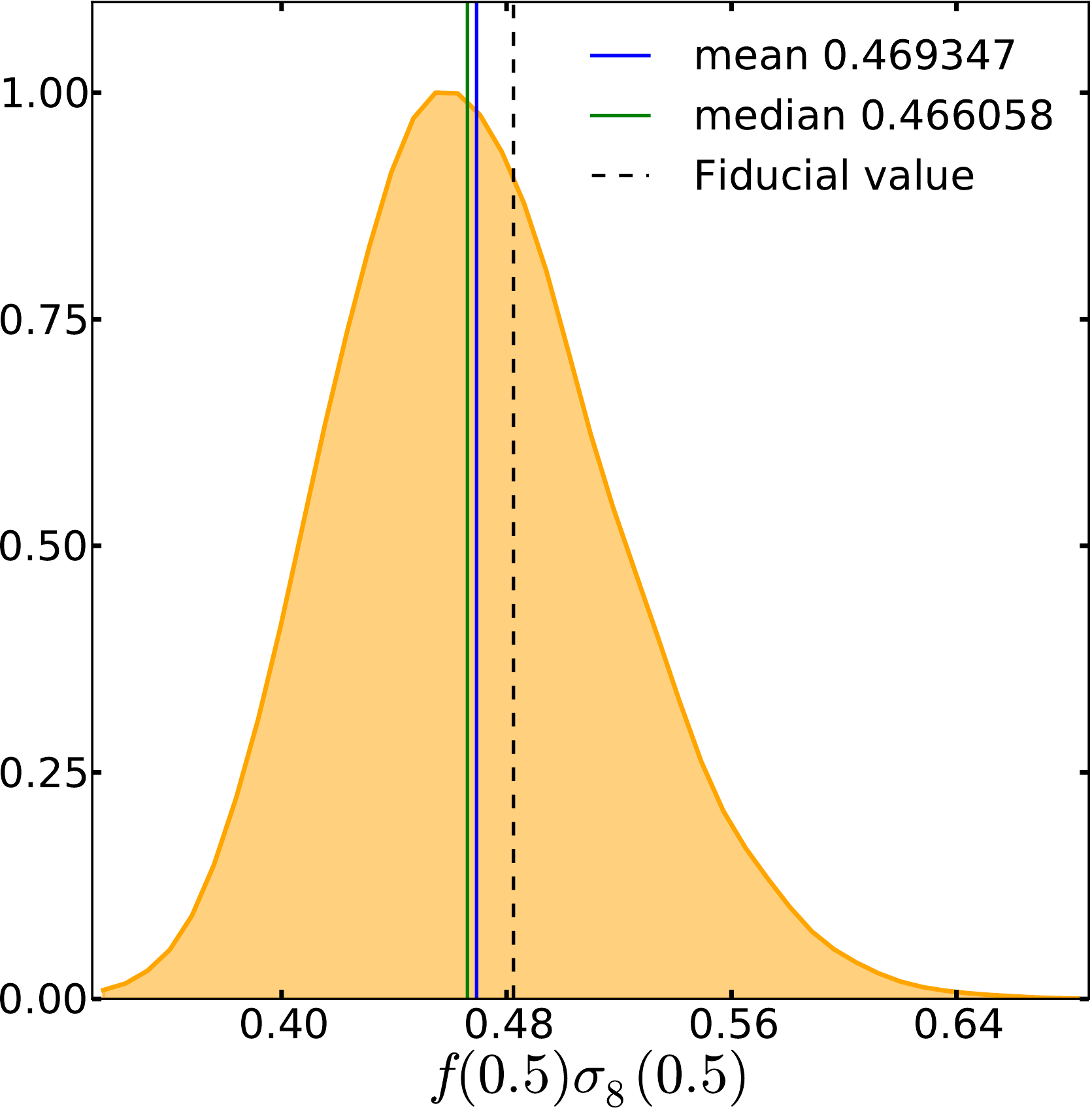} &
      \includegraphics[scale=0.315]{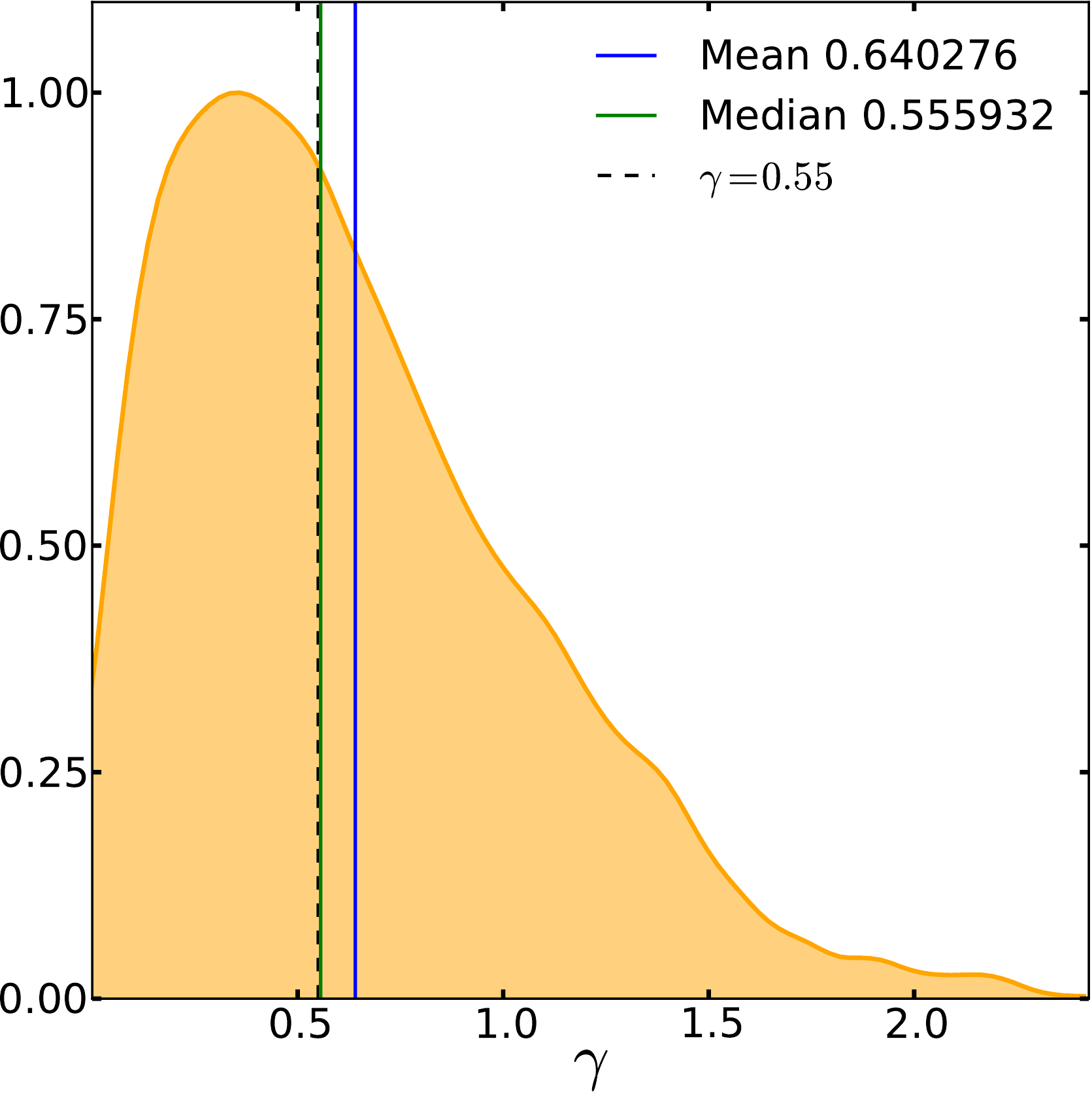}
   \end{array}$
\end{center}
   \caption{Results from the tests, described in Section \ref{sec:tests}, of our tomographic technique applied to the mean of 1000 {\sc md-patchy} mock catalogs. The left panel shows derived constraints on $D_{\rm M}(z=0.5)$ and $H(z=0.5)$ on the parameter space given in eq. (\ref{eq:paramsPatchylinb}). The central panel shows derived constrains on $f\sigma_8$ at $z=0.5$, on the same parameter space as the left panel. The right panel shows constraints on the growth index $\gamma$ on the parameter space given in eq. (\ref{eq:paramsPatchylinbgamma}).}
   \label{fig:patchymean}
\end{figure*}

We test our model for $\omega(\theta)$ and its full covariance matrix against the combined-sample {\sc md-patchy} mock catalogues. We measure the angular clustering using the binning scheme described in Section \ref{sec:opti}, and perform fits to the mean of 1000 realisations and to a subsample of 100 realisations individually. Through MCMC analysis, we explore four parameter spaces that are extensions of the standard $\Lambda$CDM model, allowing for curvature and a free dark energy equation of state parameter, $w_{\rm{DE}}$, constant in time; keeping the spectral index $n_{\rm{s}}$ and the baryon fraction $f_{\rm{b}}$ fixed to their fiducial value. 

The first parameter space consists of
\begin{equation}
   \mathbb{P}_1 = \lbrace \Omega_{\rm K}, \Omega_\Lambda h^2, w_{\rm{DE}}, \ln(10^{10}A_{\rm{s}}), b_1, b_2, \gamma_3^-, a_{\rm{vir}} \rbrace,
   \label{eq:paramsPatchy}
\end{equation}
using the F96 bias-model in equation (\ref{eq:biasFry}), and the CGC bias-model in equation (\ref{eq:biasCGC}). The second parameter space is given by
\begin{equation} 
\mathbb{P}_2=\mathbb{P}_1\cup\lbrace b^\prime\rbrace,
   \label{eq:paramsPatchylinb}
\end{equation}
using the redshift evolution of the linear galaxy bias as in equation (\ref{eq:linb}). The other two parameter spaces are defined as
\begin{equation} 
\mathbb{P}_3=\mathbb{P}_1\cup\lbrace \gamma \rbrace,
   \label{eq:paramsPatchygamma}
\end{equation}
\begin{equation} 
\mathbb{P}_4=\mathbb{P}_2\cup\lbrace \gamma \rbrace,
   \label{eq:paramsPatchylinbgamma}
\end{equation}
where $\gamma$ is the growth index, such that the growth rate factor, $f=\frac{\partial\ln D}{\partial\ln a}$, is approximated by \citep{Linder:2005aa}
\begin{equation}
	f(a)\approx\Omega_{\rm m}^\gamma(a),
	\label{eq:fgamma}
\end{equation}
and consequently the linear growth factor is
\begin{equation}
	\ln D(a) \approx \int^a_{a_0}\frac{da}{a}\Omega^\gamma_{\rm m}(a),
	\label{eq:Dgamma}
\end{equation}
imposing the border condition $\frac{D(a_0)}{a_0}= 1$ at some $a_0$ in the matter-dominated epoch. The value of $\gamma=0.55$ recovers the predictions of General Relativity (GR) for $D(a)$ and $f(a)$, and any deviation from it (in the real data) would suggest that the clustering measurements are in tension with GR. 
We assume a Gaussian likelihood function of the form $\mathcal{L}(\mathbf{P})\propto\exp\left(-\chi^2(\mathbf{P})/2\right)$, where
\begin{equation}
 	\chi^2\left(\mathbf{P}\right) = \left[\mathbf{m}\left(\mathbf{P}\right)-\mathbf{d}\right]^T\mathbf{\sf Cov}^{-1}\left[\mathbf{m}\left(\mathbf{P}\right)-\mathbf{d}\right],
 \label{eq:chi2}
\end{equation}
$\mathbf{P}$ is a vector with the parameter values, $\mathbf{d}$ is the full data vector containing all the measurements of $\omega^{(p,q)}(\theta)$, $\mathbf{m}\left(\mathbf{P}\right)$ is the model vector given $\mathbf{P}$,  and $\mathbf{\sf Cov}$ is the full covariance matrix described in Section \ref{sec:covmod}.

For each test we derive values of $D_{{\rm{M}}}(z_{\rm{ref}})$, $H(z_{\rm{ref}})$, $f(z_{\rm{ref}})$ and $\sigma_8(z_{\rm{ref}})$ from the cosmological parameters, at the reference redshift $z_{\rm{ref}}=0.5$. These quantities are more familiar in galaxy clustering analyses, and easier to refer to. We emphasise though, that these are derived quantities, and we are not measuring them at that particular redshift, but rather constraining the cosmological parameters through the full shape of $\omega(\theta)$ and its redshift evolution.

We performed tests constraining $\mathbb{P}_1$ using F96 and $\mathbb{P}_2$ for different minimum angular scales, $\theta_{\rm{min}}(\bar z^{(p,q)})$, using the mean of the mocks. We find that using smaller angular scales than $\theta_{\rm{min}}(\bar z^{(p,q)})=20{\rm{Mpc}}/h$ (in the BOSS fiducial cosmology) results in biased constraints, while larger values only increase the errors without changing the mean. In the rest of this analysis, we use this minimum scale.

The CGC model for the galaxy-bias evolution, given by equation (\ref{eq:biasCGC}), does not describe $b(z)$ of the mock catalogs, resulting in biased constraints of $\gtrsim1\sigma$ in all the tests.

Figure \ref{fig:patchymean} shows the results obtained using the mean of the mocks for different tests. The left panel shows constraints on $D_{{\rm{M}}}(z_{\rm{ref}})$ and $H(z_{\rm{ref}})$ on $\mathbb{P}_2$, i.e. using the linear bias in equation (\ref{eq:linb}). We do not see any significant deviation in this case, finding $0.1\sigma$ and $0.3\sigma$ for  $D_{{\rm{M}}}(z_{\rm{ref}})$ and $H(z_{\rm{ref}})$ respectively. These deviations are somewhat smaller, and the errors tighter, in the test on $\mathbb{P}_1$ using the F96 bias model in equation (\ref{eq:biasFry}). The middle panel shows constraints on $f(z_{\rm{ref}})\sigma_8(z_{\rm{ref}})$ on $\mathbb{P}_2$, and the right panel shows the constraints on the growth index $\gamma$ on $\mathbb{P}_4$. In these two cases, the results on $\mathbb{P}_1$ and $\mathbb{P}_3$, using the F96 model, are also unbiased and the errors smaller. In all three panels, the fiducial values, shown by the dashed lines, are those corresponding to the true cosmology of the {\sc md-patchy} mock catalogues.

\begin{figure}
   \includegraphics[scale=0.4]{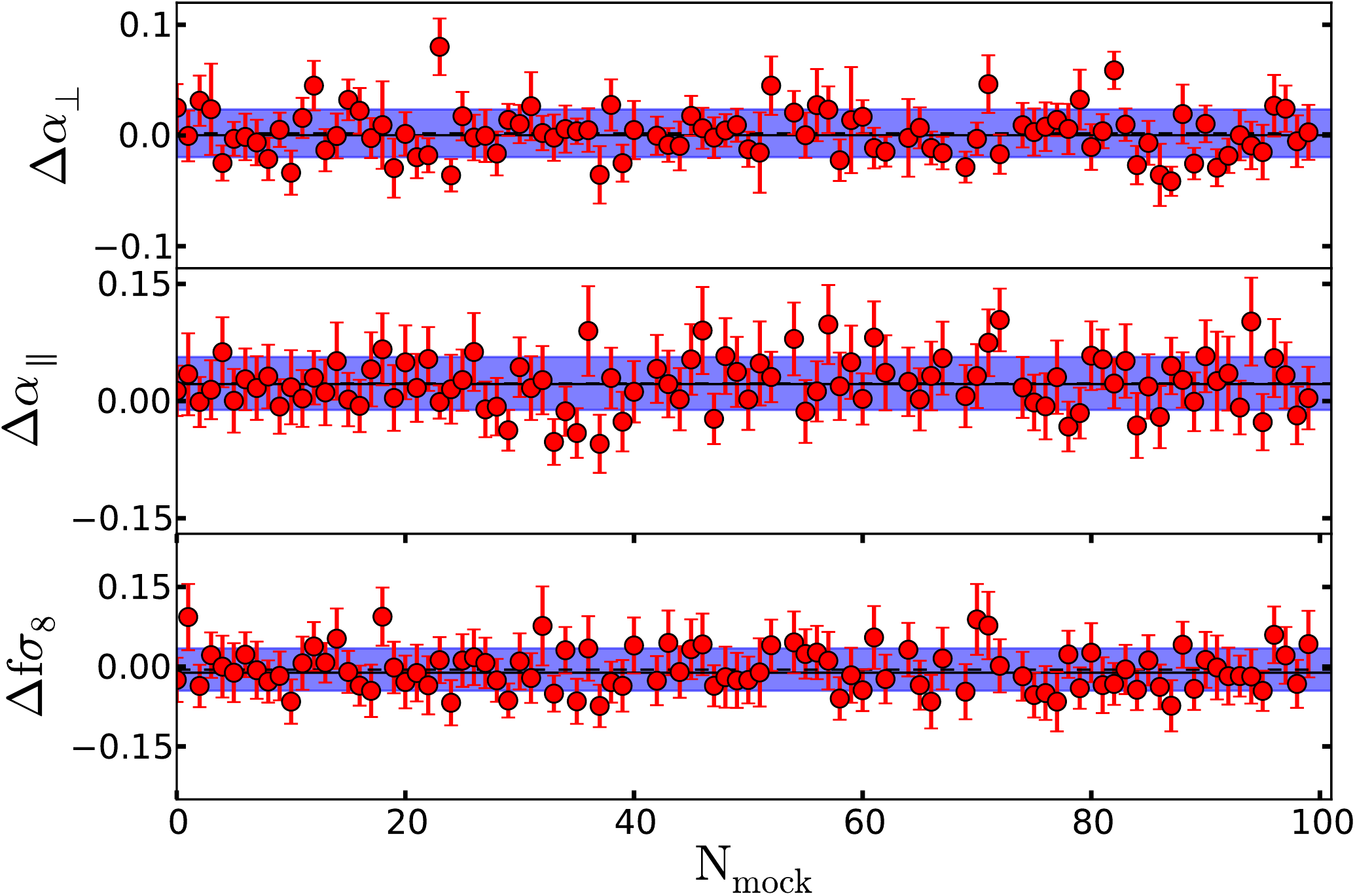} 
   \caption{Deviations between the true and the obtained values for the derived parameters $\alpha_\perp$, $\alpha_{||}$ and $f\sigma_8$ at $z=0.5$, from the individual fits (symbols) on a subset of 100 {\sc md-patchy} mock catalogues. Error bars correspond to the estimated error on each fit, while the blue bands show the sample standard deviation. The upper panel shows the deviations on $\alpha_\perp$, the middle panel shows the deviations on $\alpha_{||}$, and the lower panel shows those of $f\sigma_8$.}
   \label{fig:patchyscatter}
\end{figure}

Figure \ref{fig:patchyscatter} shows the results of the same test, this time fitting the subset of 100 mocks individually, constraining $\mathbb{P}_2$. The upper panel shows the deviations from the true values on 
\begin{equation}
    \alpha_\perp = \frac{D_{\rm M}(z)r^{\rm{fid}}_{\rm s}(z_{\rm d})}{D^{\rm{fid}}_{\rm M}(z)r_{\rm s}(z_{\rm d})},
    \label{eq:alphaperp}
\end{equation}
the middle panel shows those of
\begin{equation}
    \alpha_{||} = \frac{H^{\rm{fid}}(z)r_{\rm s}(z_{\rm d})}{H(z)r^{\rm{fid}}_{\rm s}(z_{\rm d})},
    \label{eq:alphapar}
\end{equation}
and the lower panel the deviations on $f\sigma_8$ at $z_{\rm{ref}}$, where $r_{\rm s}(z_{\rm d})$ is the sound horizon at the drag redshift, and ``$\rm{fid}$'' stands for the fiducial values in the mock's cosmology. The error bars correspond to the error from the individual fits, and the blue band corresponds to the standard deviation of the sample. The solid and dashed lines are the median and the mean of the distribution respectively, which are practically indistinguishable because the individual values are normally distributed.

Overall, these tests show that, through the redshift evolution of the full shape of $\omega(\theta)$, we can recover an expansion history and RSD information that is in very good agreement with the fiducial cosmology of the mocks, with the $0.3\sigma$ deviation in $H(z)$ being the largest one. These tests also confirm the importance of a sensible choice of a model for the galaxy-bias evolution \citep[see e.g.][]{Clerkin:2015aa}, and show that our simple linear model in eq. (\ref{eq:linb}) is flexible enough for the description of the redshift evolution of the linear bias of the BOSS galaxy sample.


\section{The linear bias of the BOSS galaxy sample}\label{sec:bias}

Assuming the best-fitting $\Lambda$CDM cosmology from Planck, we measure the linear galaxy bias in each redshift shell in two ways. First, we fit all auto correlations independently (shell by shell), fitting $b_1$ and marginalising over  $b_2$ and $\sigma_8$, the amplitude of (linear-theory) density fluctuations in spheres of $R=8 \ \rm{Mpc}/h$. We impose a prior on $\sigma_8$ from Planck. Secondly, we fit all redshift shells simultaneously, using each of our three models for $b(z)$ (Linear, F96 and CGC), and marginalising over the other three nuisance parameters of our model for $\omega(\theta)$. For comparison, we repeated the first test on the mean of the {\sc md-patchy} mocks, using the correct $P_L(k)$ and $\sigma_8$ for the mocks cosmology.

\subsection{The redshift evolution of the linear bias of BOSS galaxies}

None of the models for the redshift evolution of the linear galaxy-bias used in this analysis is able to simultaneously fit, within the errors, the first $16$ measurements and the two high-redshift ones. A possible explanation for this is that, above $z\gtrsim0.6$, the BOSS galaxy sample behaves as a flux-limited one \citep[see e.g.][]{Saito:2016aa}, i.e. only intrinsically bright galaxies can be observed at those distances, while intrinsically fainter ones are not in the sample. On the other hand, at $z\lesssim0.6$, this galaxy sample is much closer to a volume-limited sample, thus practically all galaxies brighter than a certain absolute magnitude $M_{\rm{lim}}$ have been observed. In practice, this means that above $z\gtrsim0.6$, the effective clustering amplitude is not representative of a given galaxy population, but rather dominated by observational systematics. This effect has not been observed before in other clustering analyses of BOSS galaxies in redshift bins \citep{Reid:2014aa, Saito:2016aa}, because the binning in those analyses consisted in much wider redshift-bins, hindering this variation in the amplitude of the clustering signal.

Not being able to reproduce the linear bias, hence the clustering amplitude of these high-redshift measurements, has two important consequences. An incorrect estimation of the linear galaxy bias, for a given redshift shell, implies that all estimates of the covariance in equation (\ref{eq:covtheta}) including this redshift shell will be incorrect. Secondly, the F96 and CGC models depend on the growth factor $D(z)$, which encodes cosmological information. Then, non-cosmological variations in the linear galaxy-bias could result in biased cosmological constraints. For this reason, and in order to be conservative, we exclude the galaxies above redshift $z=0.6$ from the rest of the analysis. This means that we do not use the last three high-redshift bins, even though the $16$th shell at $z\sim0.6$ seems to be within the errors.

\begin{figure}
   \includegraphics[scale=0.45]{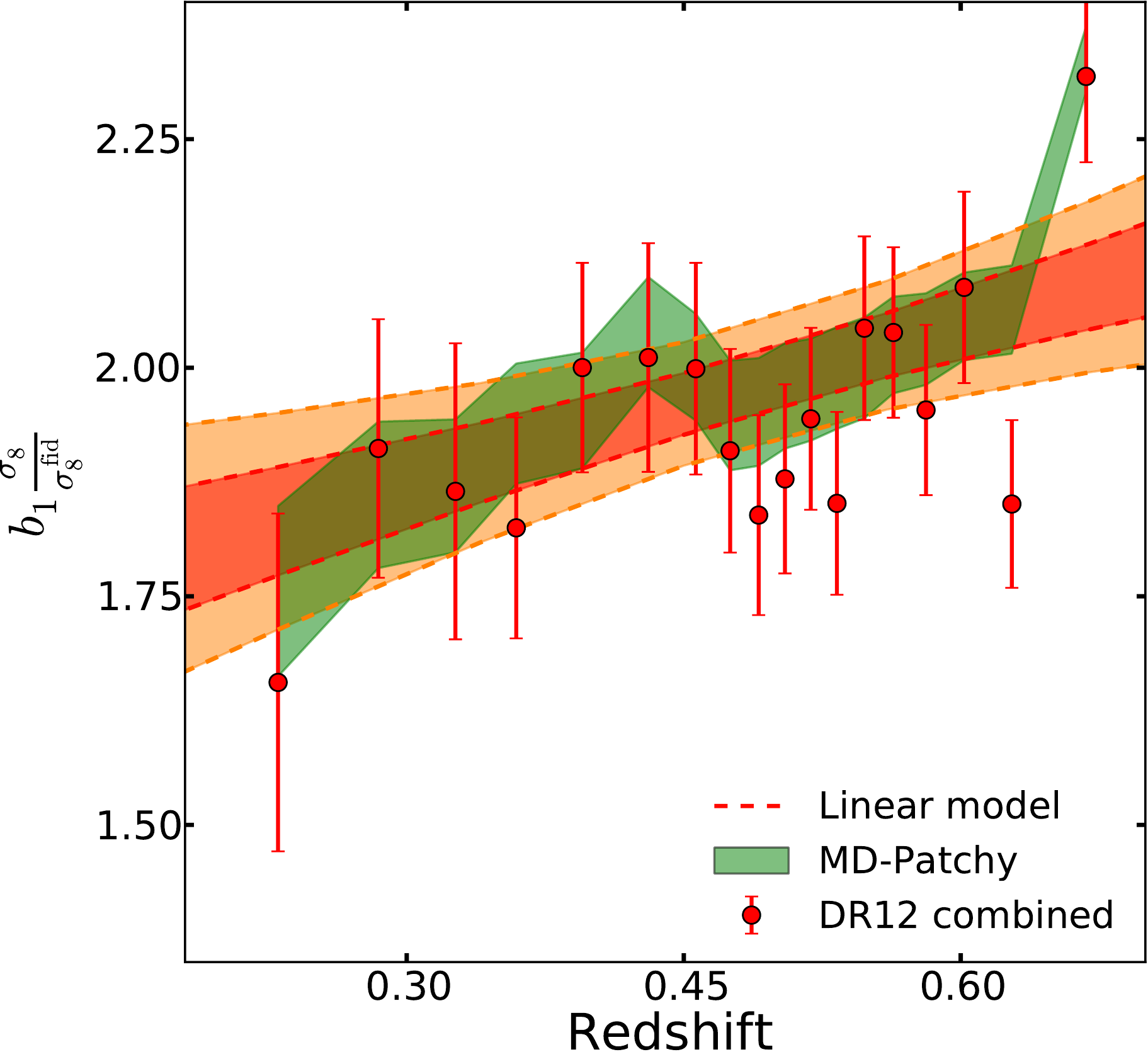}
   \caption{Redshift evolution of the linear galaxy bias. Red symbols show individual fits to 18 $\omega(\theta)$ measurements on BOSS. The green band shows the result of performing the same exercise on the mean of the {\sc md-patchy} mock catalogues. The dashed lines show the $68\%$ and $95\%$ confidence intervals obtained by fitting all clustering measurements simultaneously (excluding the three highest-redshift ones) with the bias model given in equation (\ref{eq:linb}).}
   \label{fig:dr12bias}
\end{figure}

Figure \ref{fig:dr12bias} shows the measured linear galaxy-bias normalised by the ratio of the corresponding $\sigma_8$ of each cosmology and the fiducial one coming from the Planck prior. The individual measurements are shown by the red circles, where error bars correspond to the $1\sigma$ marginalised error. The joint fit assuming the linear galaxy-bias evolution of equation (\ref{eq:linb}) is shown by the dashed lines, where the different levels correspond to one and two $\sigma$ confidence levels. We exclude the last three high-redshift measurements from this fit. The green band shows the $1\sigma$ region of the individual fits on the mean of the mock catalogues.

\subsection{The impact of the bias redshift evolution of BOSS galaxies on cosmological constraints}

We test the impact that assuming any of the three models for the redshift evolution of the linear galaxy-bias has on the obtained cosmological constraints. For this we combine our measurements of the full shape of $\omega(\theta)$ with Planck, and perform an MCMC analysis. Using each of the three models, we explore an extension of the standard $\Lambda$CDM model, allowing for the dark energy equation-of-state parameter, $w$, assumed to be constant in time, to deviate from the canonical value of $-1$. The basic cosmological parameters explored are listed in the first block of table \ref{tab:paramsMCMC}.

\begin{figure}
   \includegraphics[scale=0.335]{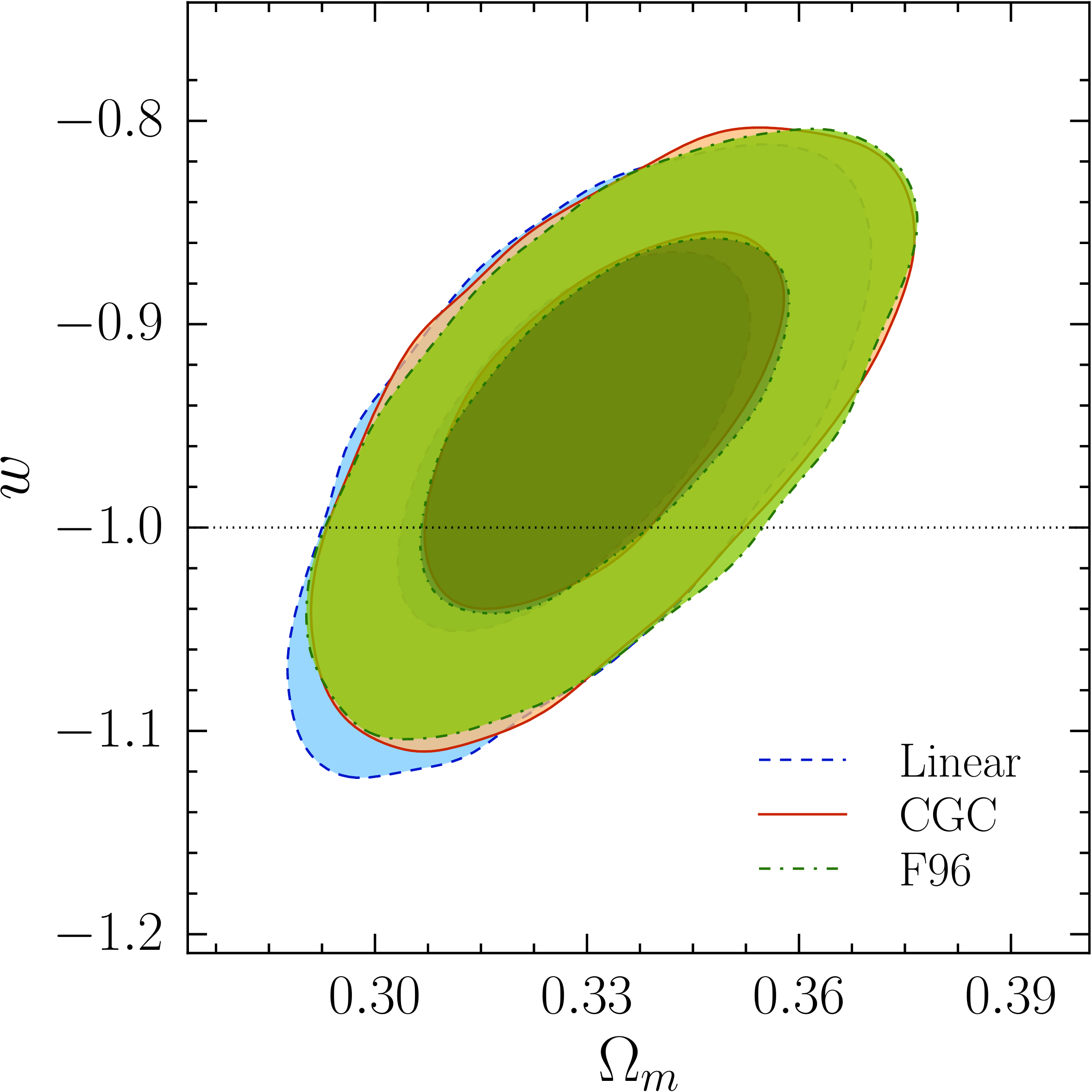}
   \caption{Cosmological constraints obtained from the ``$\rm{Planck}+\omega(\theta)$'' combination using each of our three models for the redshift-evolution of the linear galaxy-bias. Contours show the $68\%$ and $95\%$ confidence intervals on the $\Omega_{\rm m}-w$ plane.}
   \label{fig:Omw3bias}
\end{figure}

Figure \ref{fig:Omw3bias} shows the constraints on the total mass density parameter, $\Omega_{\rm m}$, and $w$, obtained from the ``$\rm{Planck}+\omega(\theta)$'' combination. The blue dashed line corresponds to the use of the linear model for $b_1(z)$, the red solid line to CGC, and the green dash-dotted line corresponds to the F96 bias model. Unlike what we find in the tests on the mock catalogues in Section \ref{sec:tests}, where different assumptions for the evolution of the linear galaxy-bias result in differences in the final cosmological constraints, the ``$\rm{Planck}+\omega(\theta)$'' combination seems to be robust against the different assumptions within the errors. The three mean values recovered in each case are within $0.16\sigma_{\rm{Linear}}$ from the linear bias case and, in both the CGC and the F96 cases, the errors are only about $4\%$ tighter compared to the linear case. Mean values and confidence intervals for the linear case are shown in Section \ref{sec:wcdm}. Our interpretation is that, firstly, the inclusion of CMB data breaks degeneracies within parameters that are present in the $\omega(\theta)$-only likelihood function, which could solve the $1\sigma$ deviation from the CGC model (assuming that the bias evolution of the mocks represents well that of the data). Secondly, the assumed models for the redshift evolution of the linear galaxy-bias are well motivated on the characteristics of BOSS galaxies (see Section \ref{sec:grptBOSS}), thus large deviations are not expected.

\section{Cosmological constraints}\label{sec:constraints}

In this section we present constraints on cosmological parameters for the standard $\Lambda$CDM model, as well as for eight different extensions described in the following subsections. For this purpose we use the July 2015 version of the publicly-available MCMC-code {\sc cosmomc} \citep{Lewis:2002aa}, modified to compute the model for $\omega(\theta)$, including non-linearities, bias and redshift-space distortions, described in Sections \ref{sec:acfmod} and \ref{sec:grptBOSS}. Although we found in the previous section that, after combining our angular clustering measurements with Planck, the different assumptions for the redshift evolution of the linear galaxy-bias do not have a significant impact on the cosmological constraints, here we take a conservative approach and only use the linear model in equation (\ref{eq:linb}).

\begin{table*}
\caption{Summary of the cosmological parameters explored in this analysis. Basic $\Lambda$CDM parameters are in the first block, while those of extended cosmological models are listed in the second block. The last block shows derived parameters quoted in avery case.}
\begin{center}
\begin{tabular}{c c c l}
\hline\hline
Parameter & Range & Fiducial value & Description\\ \hline
$\Omega_b h^2          $ & $[0.005,0.1]$ & - & Physical baryon density \\
$\Omega_c h^2          $ & $[0.001,0.99]$ & - & Physical CDM density\\ 
$100\theta_{\rm{MC}}     $ & $[0.5,10]$ & - &  Approximate angular size of $r_{\rm s}$ at recombination$^a$\\
$\tau                            $ & $[0.01,0.8]$ & - & Optical depth to the reionisation epoch \\ 
$\ln(10^{10}A_s)         $ & $[2,4]$ & - & Scalar spectral amplitude$^b$\\ 
$n_s                   $ & $[0.8,1.2]$ & - & Scalar spectral index$^b$\\ 

\hline
$w_0                     $ & $[-0.3,-3]$ & $-1$ & Present-day $w_{\rm{DE}}$\\ 
$w_a                     $ & $[-2,2]$ & $0$ & Time dependence of $w_{\rm{DE}}$\\ 
$\Omega_{\rm K}$ & $[-0.3,0.3]$ & $0$ & Curvature contribution to the energy density \\
$\Sigma m_\nu          $ & $[0,2]\rm{eV}$ & $0.06\rm{eV}$ & Total sum of neutrino masses \\ 
$\gamma$ & $[0,2]$ & - & Growth index \\
\hline

$H_0                  $ & $[20,100]$ & - & Hubble constant\\ 
$\Omega_{\rm m}      $ & - & - & Present-day total\\
 					& & & matter density\\ 
$\Omega_\Lambda             $ & - & - & Dark energy density \\ 
$\sigma_8             $ & - & - & Amplitude of linear-theory density fluctuations\\
				& & & in spheres of R = $8 \rm{Mpc}/h$ \\ 
${\rm{Age}}/{\rm{Gyr}}$ & - & - & Age of the Universe \\ 
\hline\hline
\end{tabular}
\label{tab:paramsMCMC}
\end{center}
\begin{flushleft}
\hskip 2.4cm $^a${\it As defined in the July 2015 version of {\sc cosmomc}}.\\
\hskip 2.4cm $^b${\it Quoted at the pivot $k_0=0.05$ $(\rm{Mpc})^{-1}$}.
\end{flushleft}
\end{table*}

Table \ref{tab:paramsMCMC} displays the cosmological parameters explored in these analyses, the ranges in which they are allowed to vary, and fiducial values in the case that a given parameter is fixed. The first block lists the basic parameters varied in all cases, corresponding to those that characterise the standard $\Lambda$CDM cosmological model. The second block in the table lists those parameters that represent extensions of the standard cosmological model explored in this analysis. The last block in table \ref{tab:paramsMCMC} displays derived parameters quoted in each case. 

As we do in Section \ref{sec:tests}, we assume Gaussian likelihood function of the form $\mathcal{L}(\mathbf{P})\propto\exp\left(-\chi^2(\mathbf{P})/2\right)$ for our clustering measurements, where $\chi^2$ is computed as in equation (\ref{eq:chi2}).

Planck CMB constraints are only shown in Figures for comparison, and we quote results for the ``$\rm{Planck}+\omega(\theta)$'' and ``$\rm{Planck}+\omega(\theta)+\rm{SNIa}$'' cases only. Summary tables are given in appendix \ref{sec:tables} for readability, and in the text we only quote values of the most relevant parameters for each cosmological model. In every case, the values and confidence intervals correspond to those obtained after marginalising over all other parameters.

\subsection{The standard $\Lambda$CDM model}\label{sec:lcdm}

\begin{figure}
   \includegraphics[scale=0.335]{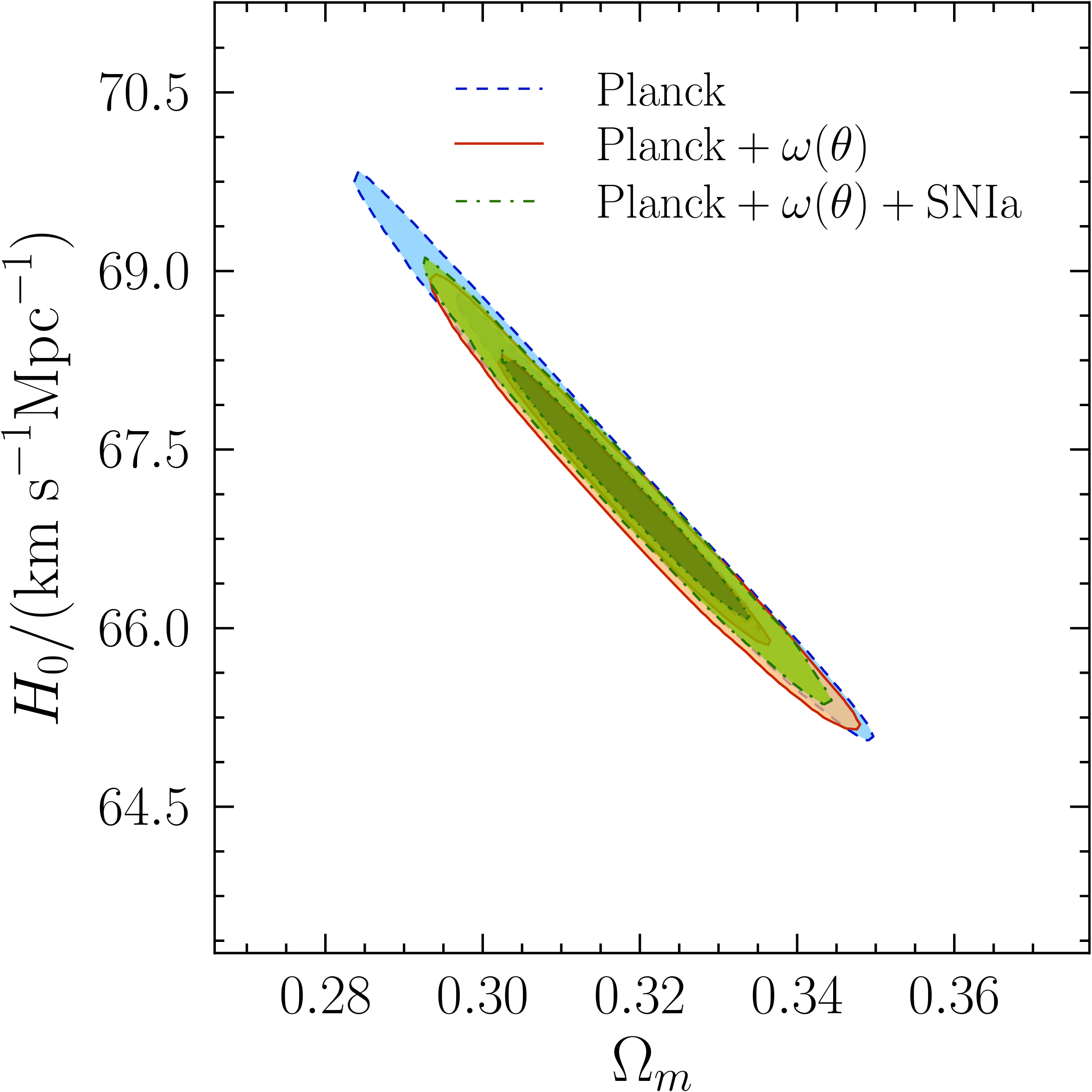}
   \caption{Marginalised $68\%$ and $95\%$ confidence interval constraints in the $\Omega_{\rm m}-H_0$ plane. The blue dashed line corresponds to $\rm{Planck}$-only constraints, the solid orange line corresponds to the constraints obtained from the $\rm{Planck}+\omega(\theta)$ combination, and the green dash-dotted line to those obtained combining $\rm{Planck}+\omega(\theta)+\rm{SNIa}$.}
   \label{fig:OmH0}
\end{figure}

We start out with the basic case: the $\Lambda$CDM model. This model has become the standard cosmological model due to its astonishing description and prediction capabilities, regarding a large list of observables.

Figure \ref{fig:OmH0} shows the marginalised 68 and 95 per cent confidence interval in the $\Omega_{\rm m}-H_0$ plane. The blue dashed line corresponds to $\rm{Planck}$-only constraints, the solid orange line shows the constraints obtained from the $\rm{Planck}+\omega(\theta)$ combination, and the green dash-dotted line shows those obtained combining $\rm{Planck}+\omega(\theta)+\rm{SNIa}$. We find that including our angular clustering measurements improves the constraints, and the subsequent addition of SNIa slightly shifts the allowed region toward higher values of $H_0$ and does not represent a significant improvement. We also find that the $\rm{Planck}+\omega(\theta)$ combination selects the highest values of $\Omega_{\rm m}$ allowed by Planck, as opposed to previous 3D clustering analyses on BOSS \citep[see e.g.][]{Sanchez:2013aa,Anderson:2014aa}. Nevertheless, our results and those mentioned are consistent within $1\sigma$. We found $\Omega_{\rm m}=0.319\pm 0.011$ for the $\rm{Planck}+\omega(\theta)$ combination, and $\Omega_{\rm m}=0.317\pm 0.011$ including SNIa. Table \ref{tab:lcdm} shows marginalised constraints for all the parameters varied in this case, as well as the derived parameters.

\subsection{The dark energy equation-of-state parameter}\label{sec:wcdm}

\begin{figure}
   \includegraphics[scale=0.335]{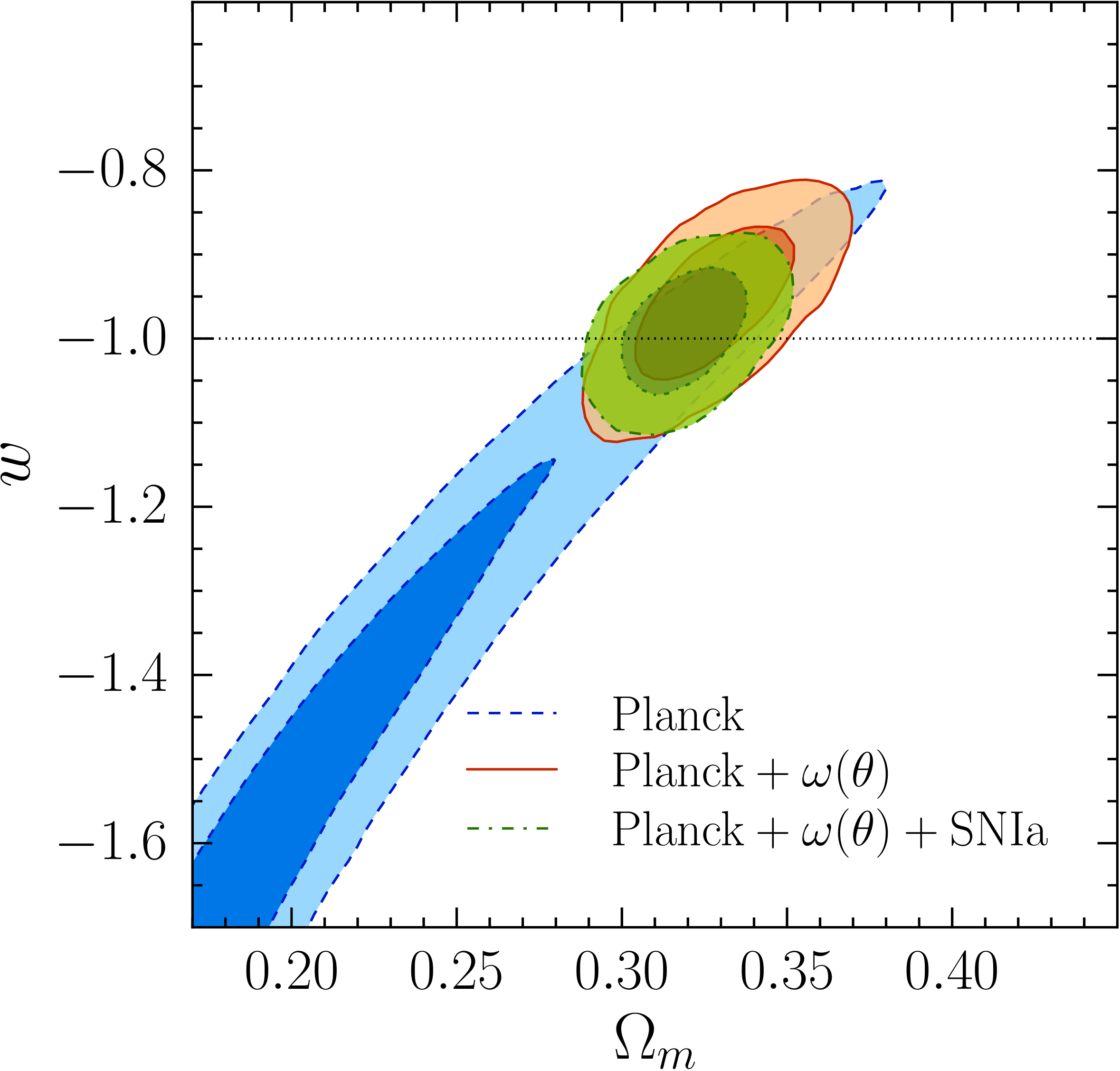}
   \caption{Marginalised $68\%$ and $95\%$ confidence interval constraints in the $\Omega_{\rm m}-w$ plane. The blue dashed line corresponds to $\rm{Planck}$-only constraints, the solid orange line corresponds to the constraints obtained from the $\rm{Planck}+\omega(\theta)$ combination, and the green dash-dotted line to those obtained combining $\rm{Planck}+\omega(\theta)+\rm{SNIa}$.}
   \label{fig:Omw}
\end{figure}

Although the standard $\Lambda$CDM model is sufficient to describe the expansion history of the Universe, as probed by the CMB power spectrum, galaxy clustering measurements and SNIa, the combination of all these observables allows us to test assumptions and generalisations of it. One of such assumptions is that the dark-energy component of the Universe is characterised by an equation of state $P_{\rm{DE}}/\rho_{\rm{DE}}\equiv w_{\rm{DE}} = -1$ constant in time. Thus the first tested extension of the standard cosmological model is to treat $w_{\rm{DE}}$ as a free parameter ($w$CDM model), assuming it is constant in time.

Figure \ref{fig:Omw} shows the marginalised 68 and 95 per cent confidence interval constraints in the $\Omega_{\rm m}-w$ plane. As before, the blue dashed line corresponds to $\rm{Planck}$-only constraints, the solid orange line to the results obtained from the $\rm{Planck}+\omega(\theta)$ combination, and the green dash-dotted line to those obtained combining $\rm{Planck}+\omega(\theta)+\rm{SNIa}$. We find that including our angular clustering measurements significantly improves the constraints obtained by Planck, where we found a value of $\Omega_{\rm m}=0.328\pm 0.016$ and $w = -0.958^{+0.063}_{-0.055}$, in very good agreement with the $\Lambda$CDM results. In this case, the $\rm{Planck}+\omega(\theta)+\rm{SNIa}$ combination improves the constraints even more, resulting in $\Omega_{\rm m}=0.319\pm 0.012$ and $w = -0.991\pm 0.046$, again in very good agreement with the $\Lambda$CDM case. A summary of the constraints obtained in this case can be found in table \ref{tab:wcdm}.

\begin{figure*}
	\begin{center}
	 $\begin{array}{cc}
		\includegraphics[scale=0.284]{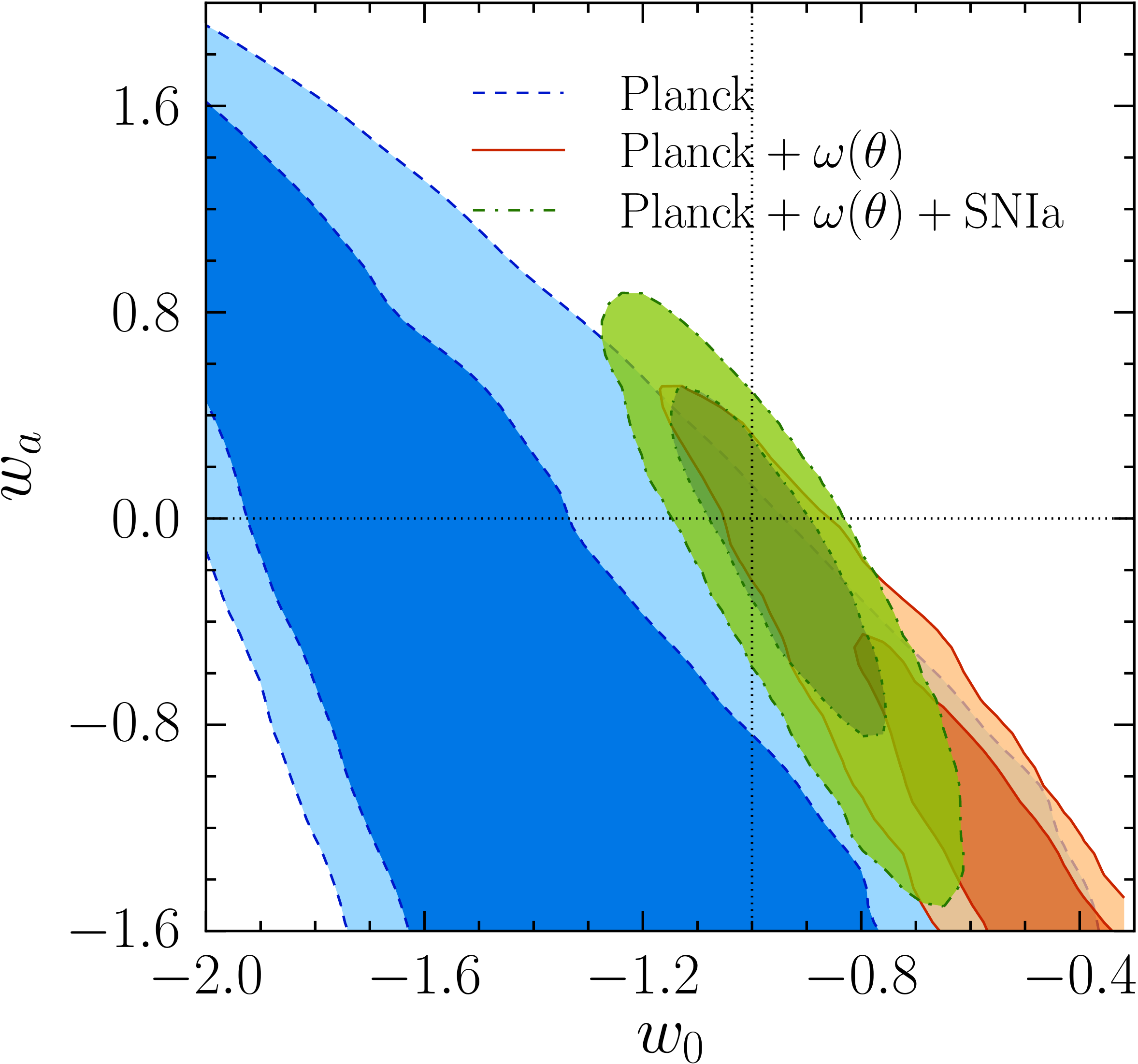} &
		\includegraphics[scale=0.405]{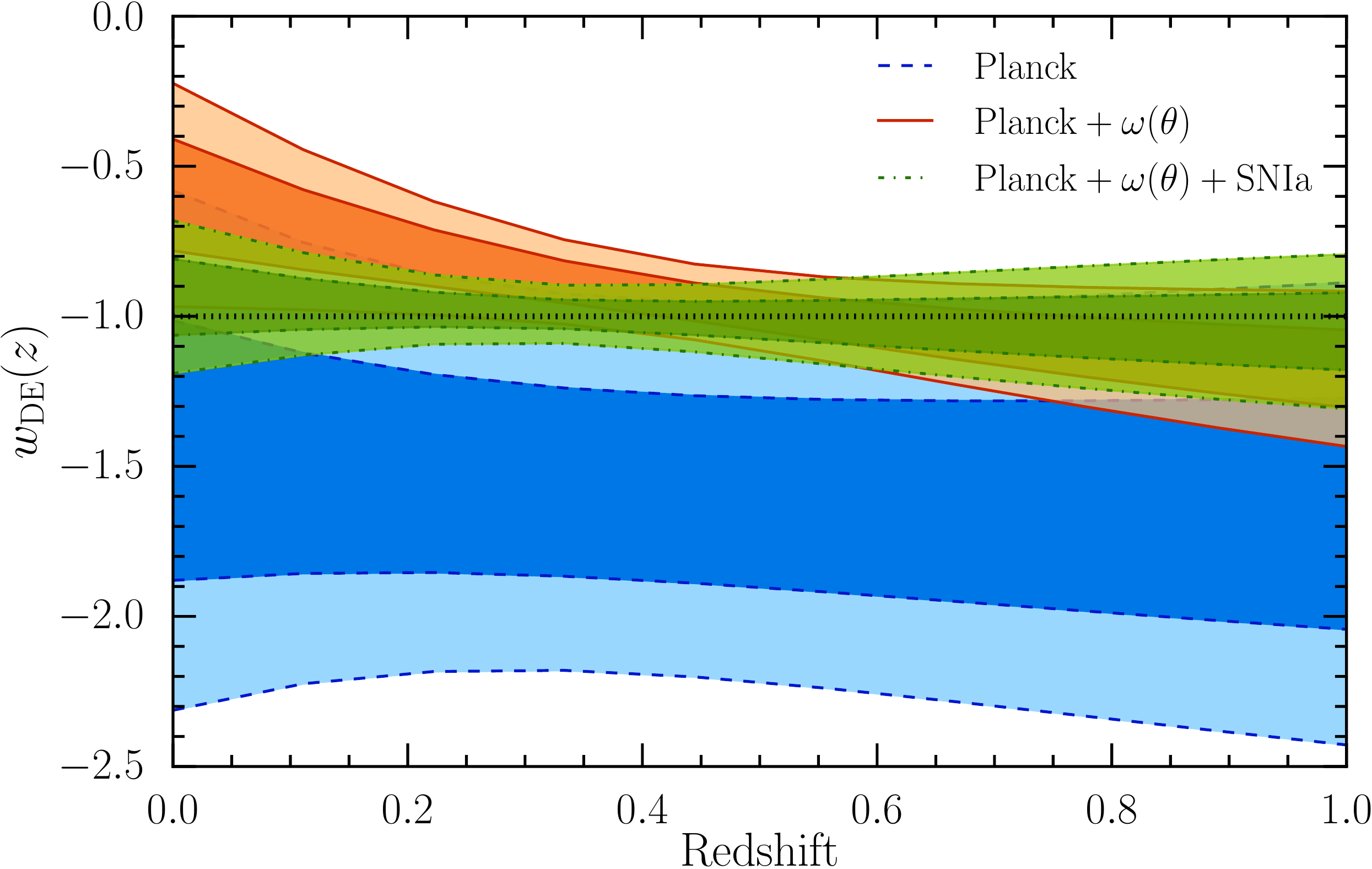}
	 \end{array}$
	\end{center}
	\caption{{\it Left}: Marginalised $68\%$ and $95\%$ confidence interval constraints in the $w_0-w_a$ plane. The blue dashed line corresponds to $\rm{Planck}$-only constraints, the solid orange line corresponds to the constraints obtained from the $\rm{Planck}+\omega(\theta)$ combination, and the green dash-dotted line to those obtained combining $\rm{Planck}+\omega(\theta)+\rm{SNIa}$. {\it Right}: Marginalised $68\%$ and $95\%$ confidence interval constraints on the redshift evolution of $w_{\rm{DE}}(z)$ using the CPL parametisation. The blue dashed line corresponds to $\rm{Planck}$-only constraints, the solid orange line corresponds to the constraints obtained from the $\rm{Planck}+\omega(\theta)$ combination, and the green dash-dotted line to those obtained combining $\rm{Planck}+\omega(\theta)+\rm{SNIa}$.}
	 \label{fig:w0wa}
\end{figure*}

Next, we allow $w_{\rm{DE}}$ to vary over time ($w_0w_a$CDM model), following the standard linear parametrisation of \cite{Chevallier:2001aa} and \cite{Linder:2003aa} (CPL), given by
\begin{equation}
 	w_{DE}(a) = w_0 + w_a(1-a). 
	\label{eq:w0wa}
\end{equation} 
The marginalised 68 and 95 per cent confidence interval constraints, in the $w_0-w_a$ plane, are shown in the left panel of Figure \ref{fig:w0wa}. In this case, we see a strong degeneracy between these two parameters for the $\rm{Planck}$ only and the $\rm{Planck}+\omega(\theta)$ combinations, where the fiducial $\Lambda$CDM values for these parameters, shown by the dotted lines, are only within the $95\%$ confidence interval, suggesting a mild tension with the standard cosmological model. Nevertheless, adding SNIa breaks this degeneracy and eliminates this tension. In this, case we find $w_0-0.94\pm 0.13$ and $w_a = -0.23^{+0.51}_{-0.42}$. Table \ref{tab:wacdm} summarises the cosmological constraints for this case.

\subsection{Non spatially-flat Universes}\label{sec:kcdm}

\begin{figure}
   \includegraphics[scale=0.335]{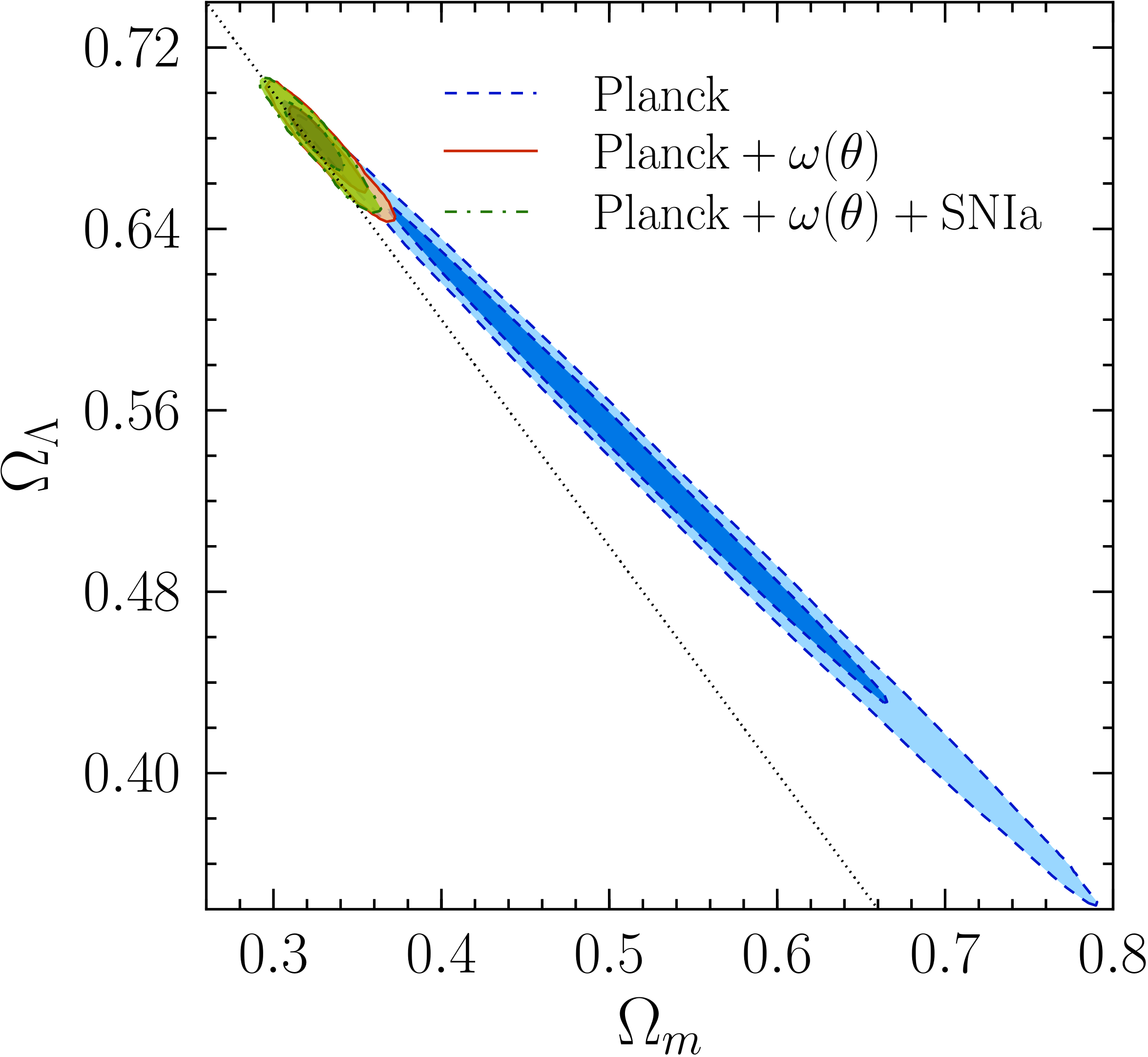}
   \caption{Marginalised $68\%$ and $95\%$ confidence interval constraints in the $\Omega_{\rm m}-\Omega_\Lambda$ plane, relaxing the flat-space condition. The blue dashed line corresponds to $\rm{Planck}$-only constraints, the solid orange line corresponds to the constraints obtained from the $\rm{Planck}+\omega(\theta)$ combination, and the green dash-dotted line to those obtained combining $\rm{Planck}+\omega(\theta)+\rm{SNIa}$.}
   \label{fig:OmOL}
\end{figure}

Another assumption of the standard $\Lambda$CDM model is that the Universe is spatially flat, which implies that its total energy density is equal to the critical one. We test this assumption of flatness by including the $\Omega_{\rm K}$ parameter.

The first case we analyse assumes $w_{\rm{DE}}\equiv-1$ ($k$CDM model). Figure \ref{fig:OmOL} shows the marginalised 68 and 95 per cent confidence interval constraints in the $\Omega_{\rm K}-\Omega_\Lambda$ plane, where the dotted diagonal line corresponds to spatially-flat Universes. It can be seen that relaxing the flat-space condition opens a large degeneracy in the CMB-only constraints, and that this degeneracy is broken adding low-redshift measurements of the expansion history of the Universe, greatly improving the constraints. For the $\rm{Planck}+\omega(\theta)$ combination we find $\Omega_{\rm m} = 0.329^{+0.014}_{-0.016}$, $\Omega_\Lambda = 0.676\pm 0.013$ and $\Omega_{\rm K} = -0.0043^{+0.0042}_{-0.0035}$, while for the full $\rm{Planck}+\omega(\theta)+\rm{SNIa}$ combination, we find $\Omega_{\rm m} = 0.324^{+0.011}_{-0.014}$, $\Omega_\Lambda = 0.679^{+0.013}_{-0.009}$ and $\Omega_{\rm K} = -0.0028\pm 0.0038$, in excellent agreement with a spatially-flat Universe, as well as with the results for the $\Lambda$CDM case. A summary of the constraints obtained in this case can be found in table \ref{tab:kcdm}.

\begin{figure}
   \includegraphics[scale=0.335]{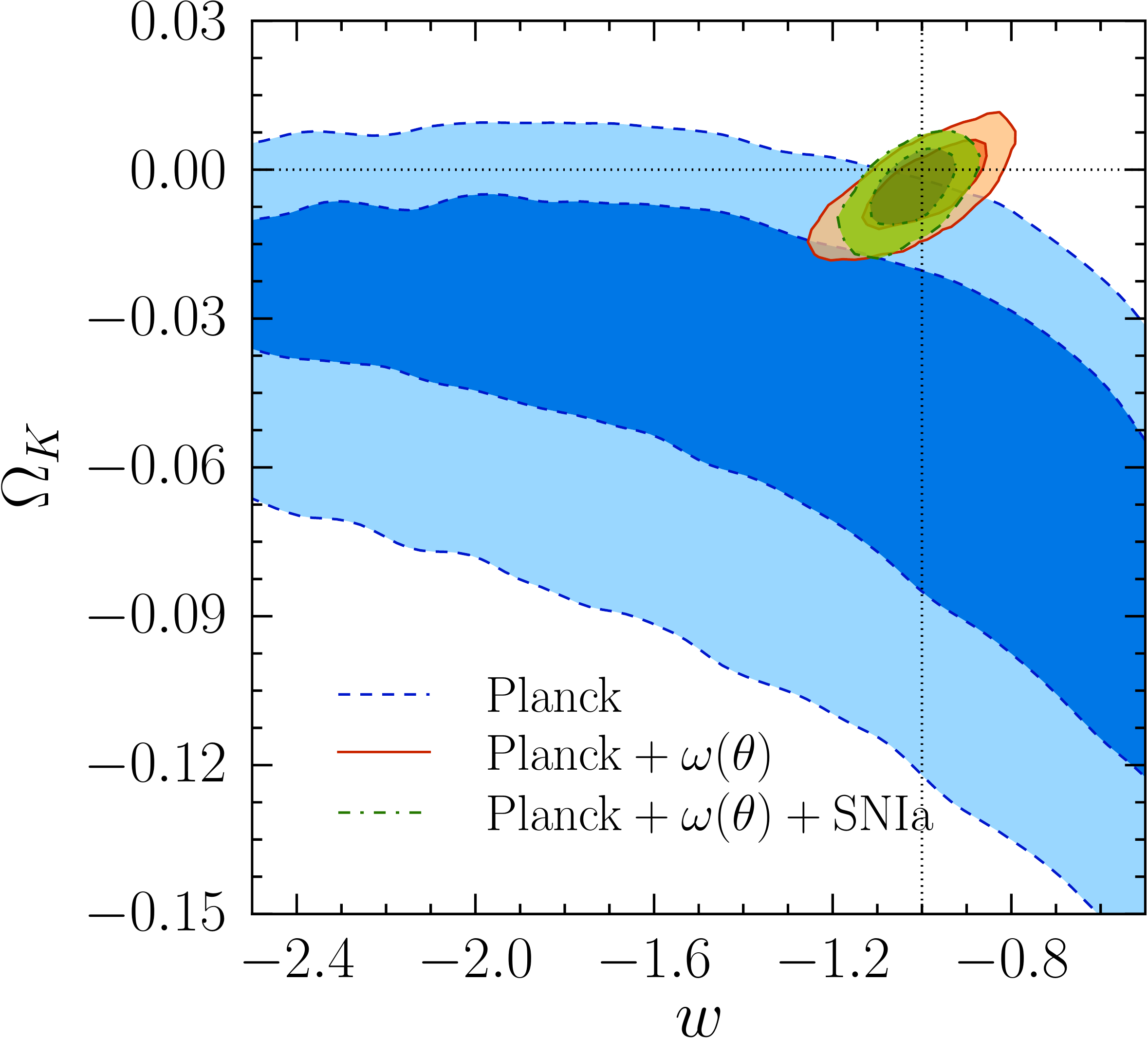}
   \caption{Marginalised $68\%$ and $95\%$ confidence interval constraints in the $w-\Omega_{\rm K}$ plane. The blue dashed line corresponds to $\rm{Planck}$-only constraints, the solid orange line corresponds to the constraints obtained from the $\rm{Planck}+\omega(\theta)$ combination, and the green dash-dotted line to those obtained combining $\rm{Planck}+\omega(\theta)+\rm{SNIa}$.}
   \label{fig:wOk}
\end{figure}

We also include $w_{\rm{DE}}$ as a free parameter in this case, assuming that its value is constant in time ($wk$CDM model). A summary of the constraints for this case can be found in table \ref{tab:wkcdm}. Figure \ref{fig:wOk} shows the marginalised 68 and 95 per cent confidence interval constraints in the $w-\Omega_{\rm K}$ plane. As always, the blue dashed line corresponds to $\rm{Planck}$-only constraints, the solid orange line corresponds to the constraints obtained from the $\rm{Planck}+\omega(\theta)$ combination, and the green dash-dotted line to those obtained combining $\rm{Planck}+\omega(\theta)+\rm{SNIa}$. Again this time, it can be seen that the inclusion of our $\omega(\theta)$ measurements on BOSS, to the CMB-only ones, significantly improves the cosmological constraints, where we find a value of $w = -1.00^{+0.10}_{-0.075}$ and $\Omega_{\rm K} = -0.0037^{+0.0057}_{-0.0051}$. Also, including SNIa further tightens the constraints, resulting in $w = -1.025^{+0.064}_{-0.055}$ and $\Omega_{\rm K} = -0.0040^{+0.0054}_{-0.0041}$, once again, in perfect agreement with the standard cosmological model.

\subsection{Massive neutrinos}\label{sec:nucdm}

Observations of neutrino oscillations (i.e., a change in neutrino flavour) imply that at least two neutrino species have non-zero mass. This is one of the most significant discoveries in the last decades, providing decisive evidence that the Standard Model (of particle physics) needs to be extended. Actually, it was for this very important discovery \citep{Fukuda:1998aa,Ahmad:2001aa,Ahmad:2002aa} that Takaaki Kajita and Arthur B. McDonald were awarded the Nobel Prize in Physics last year\footnote{``The 2015 Nobel Prize in Physics - Press Release''. Nobelprize.org. Nobel Media AB 2014. \url{www.nobelprize.org/nobel_prizes/physics/laureates/2015/press.html}}.

Although the fact that neutrinos have mass is well stablished, precise measurements of their mass is a very difficult task. The best upper limits from laboratory experiments, through tritium decay, are $m_{\nu_e} < 2$eV for electron neutrinos (see \citealt{Weinheimer:2013aa} for a review of different experiments). Nevertheless, the best constraints in their total-mass sum, including all species, comes from cosmological observations. Relic neutrinos generated in very early Universe are almost as abundant as photons, and they form what is known as the cosmic neutrino background (C$\nu$B). At the present, it is not possible to observe the C$\nu$B, but these primordial neutrinos have two important consequences for cosmology. First, they decouple from the other components before photons, free-streaming through the baryon-photon plasma and washing out small-scale anisotropies. Secondly, their mass affects the expansion rate $H$, especially at early stages. 

\begin{figure}
   \includegraphics[scale=0.335]{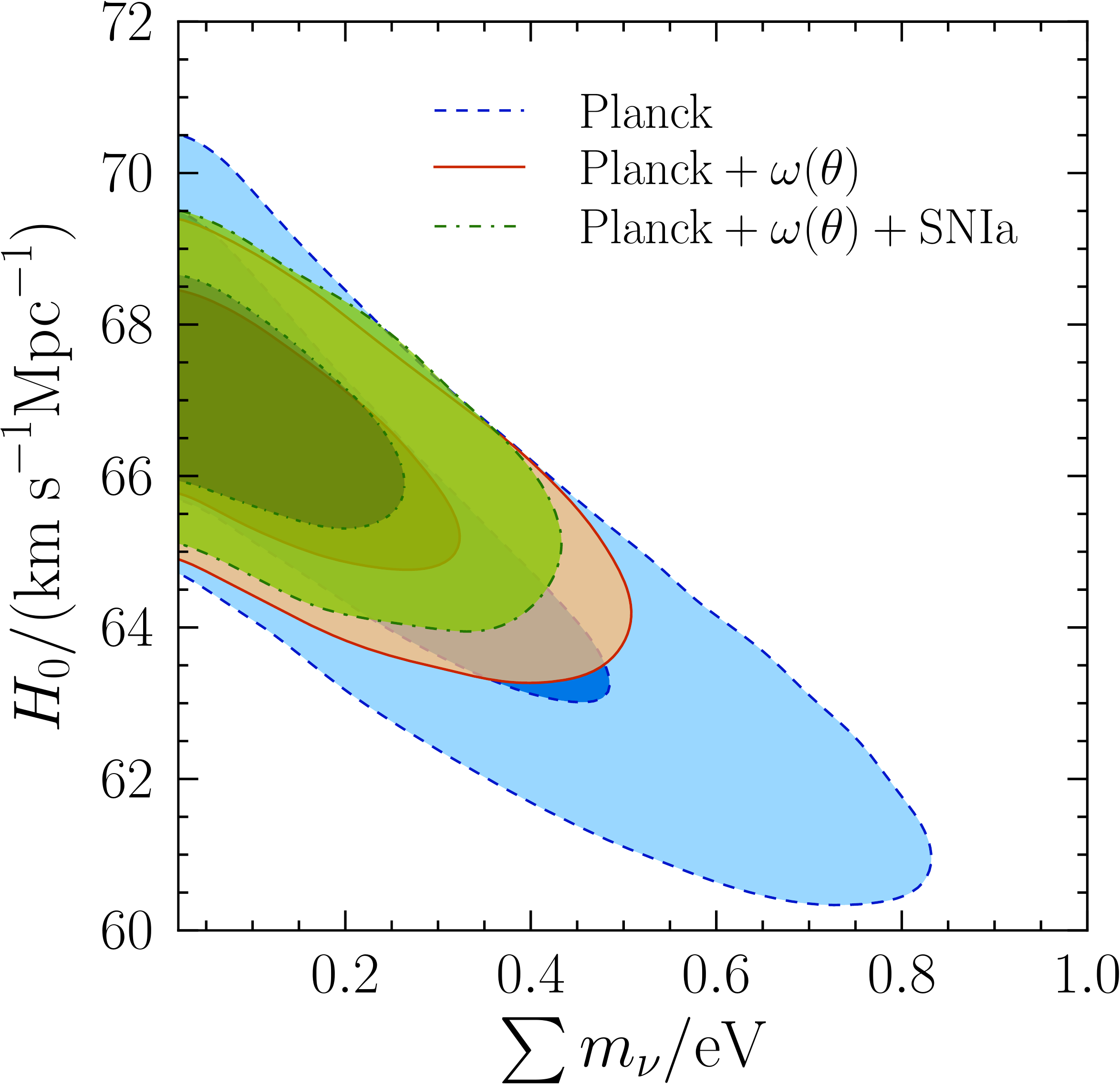}
   \caption{Marginalised $68\%$ and $95\%$ confidence interval constraints in the $\sum m_\nu/\rm{eV}-H_0$ plane. The blue dashed line corresponds to $\rm{Planck}$-only constraints, the solid orange line corresponds to the constraints obtained from the $\rm{Planck}+\omega(\theta)$ combination, and the green dash-dotted line to those obtained combining $\rm{Planck}+\omega(\theta)+\rm{SNIa}$.}
   \label{fig:mnuH0}
\end{figure}

The scales in clustering measurements affected by neutrinos are beyond what we are able to currently model, but we certainly can constrain the effect of neutrinos on the expansion rate. For this, in this section we treat the total sum of neutrino masses, $\sum m_\nu$, as a free parameter, assuming three species of equal mass. We obtain constraints within the $\Lambda$CDM and $w$CDM framework.

Figure \ref{fig:mnuH0} shows the marginalised 68 and 95 per cent confidence interval constraints in the $\sum m_\nu/\rm{eV}-H_0$ plane, fixing $w_{\rm{DE}}\equiv-1$. For the $\rm{Planck}+\omega(\theta)$ combination we find $\sum m_\nu/\rm{eV}< 0.207(0.400)$ $68\%$($95\%$) C.I. upper limits, while for the full $\rm{Planck}+\omega(\theta)+\rm{SNIa}$ combination, we find $\sum m_\nu/\rm{eV}< 0.169(0.330)$ $68\%$($95\%$) C.I. upper limits, representing one of the tightest constraints at the present (see e.g. \citealt{Neutrinos4,Neutrinos3,Neutrinos1,Neutrinos2} and our companion papers). A summary of the constraints obtained in this case can be found in table \ref{tab:mnulcdm}.

\begin{figure}
   \includegraphics[scale=0.335]{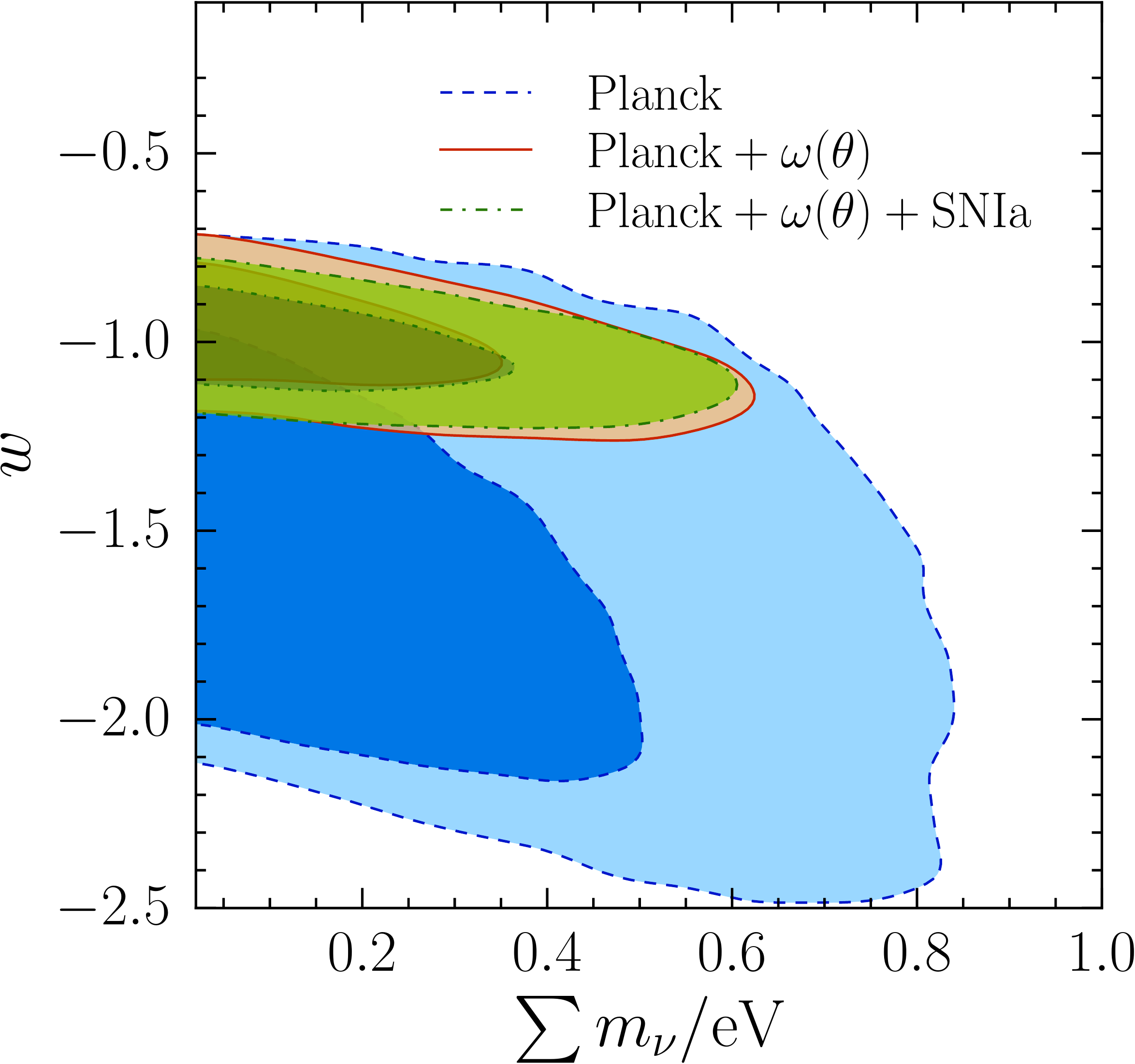}
   \caption{Marginalised $68\%$ and $95\%$ confidence interval constraints in the $\sum m_\nu/\rm{eV}-w$ plane. The blue dashed line corresponds to $\rm{Planck}$-only constraints, the solid orange line corresponds to the constraints obtained from the $\rm{Planck}+\omega(\theta)$ combination, and the green dash-dotted line to those obtained combining $\rm{Planck}+\omega(\theta)+\rm{SNIa}$.}
   \label{fig:mnuw}
\end{figure}

The results of also treating $w_{\rm{DE}}$ as a free parameter are shown in table \ref{tab:mnuwcdm}. Figure \ref{fig:mnuw} shows the marginalised 68 and 95 per cent confidence interval constraints in the $\sum m_\nu/\rm{eV}$$-w$ plane. In this case, for the $\rm{Planck}+\omega(\theta)$ combination we find $\sum m_\nu/\rm{eV}< 0.221(0.486)$ $68\%$($95\%$) C.I. upper limits, while for the full $\rm{Planck}+\omega(\theta)+\rm{SNIa}$ combination, we find $\sum m_\nu/\rm{eV}< 0.229(0.474)$ $68\%$($95\%$) C.I. upper limits. Note that the inclusion of SNIa increases the $68\%$ C.I. upper limit, decreasing the $95\%$ C.I. one, marginally suggesting non-zero masses, although we cannot claim a detection. Also, including $\sum m_\nu$ as a free parameter does not significantly degrade our constraints in $w$, resulting in $w=-1.023^{+0.063}_{-0.053}$ for the full $\rm{Planck}+\omega(\theta)+\rm{SNIa}$ combination.

\subsection{Deviations from General Relativity}\label{sec:gcdm}

The last assumption of the $\Lambda$CDM model that we test in this analysis is that of space-time being described by the theory of General Relativity. A thorough analysis of different theories beyond GR requires modifications to our methodology, such as the way the expansion history of the Universe is parametrised, which is out of the scope of this work. However, we perform a simple {\it null test}, following the parametrisation for linear perturbation growth of  \cite{Linder:2005aa}, which is decoupled from the expansion history. To a sub per-cent accuracy, the growth rate $f\equiv\frac{\partial\ln D}{\partial\ln a}$ can be approximated as in equation (\ref{eq:fgamma}), where a value of 
\begin{equation}
	\gamma = 0.55+0.05(1+w_{\rm{DE}}(z=1)), 
	\label{eq:gammaGR}
\end{equation}
for the growth index parameter recovers the prediction of GR. Thus, any deviation from this value, treating $\gamma$ as a free parameter, would suggest that general relativity should be revised.

\begin{figure}
   \includegraphics[scale=0.335]{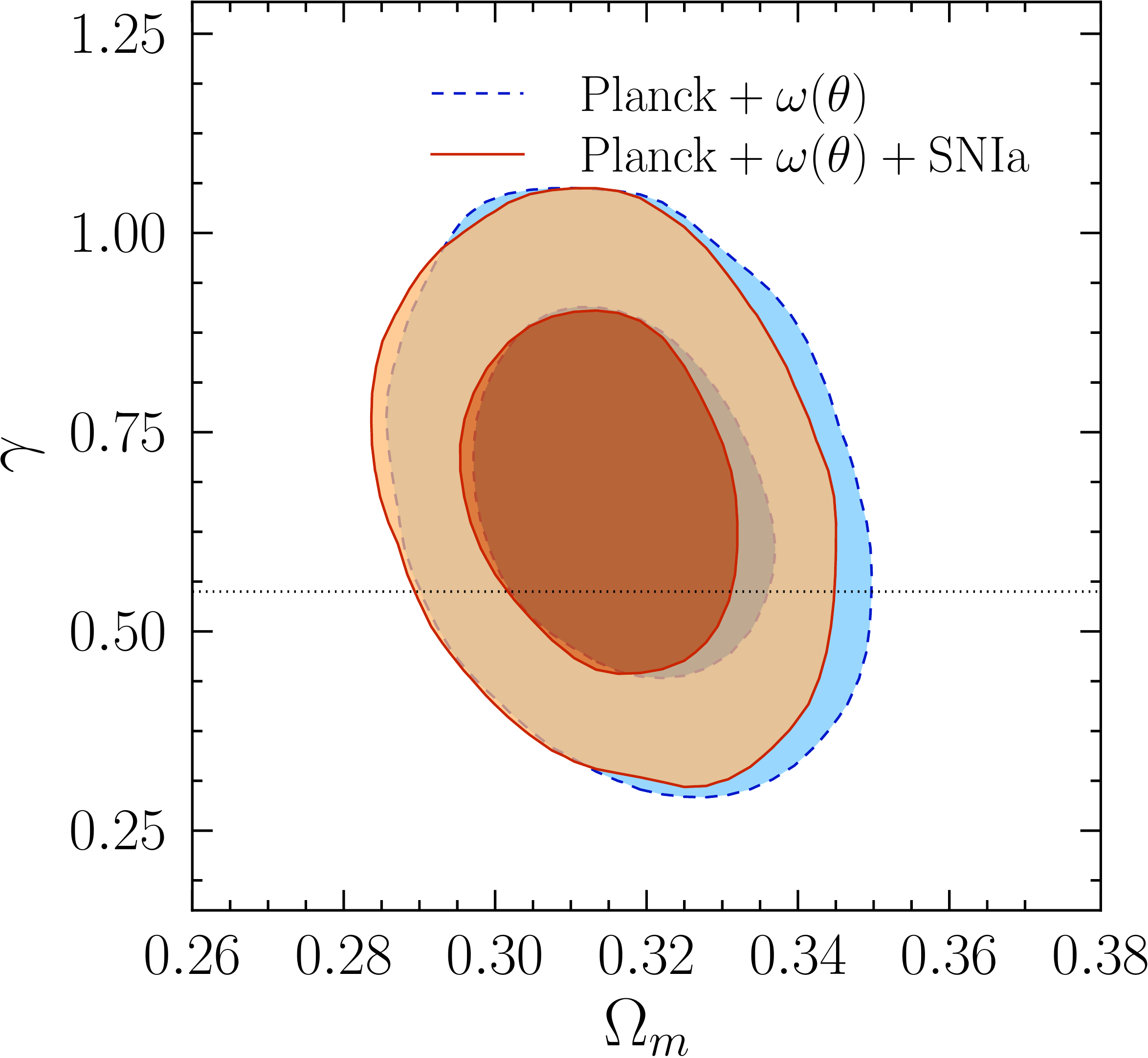}
   \caption{Marginalised $68\%$ and $95\%$ confidence interval constraints in the $\Omega_{\rm m}-\gamma$ plane. The blue dashed line corresponds to the constraints obtained by the $\rm{Planck}+\omega(\theta)$ combination, and the solid orange line line to those obtained combining $\rm{Planck}+\omega(\theta)+\rm{SNIa}$. The dotted line shows the value of $\gamma$ that recovers the GR prediction for the growth rate $f$, following equation (\ref{eq:gammaGR}).}
   \label{fig:Omgamma}
\end{figure}

\begin{figure}
   \includegraphics[scale=0.335]{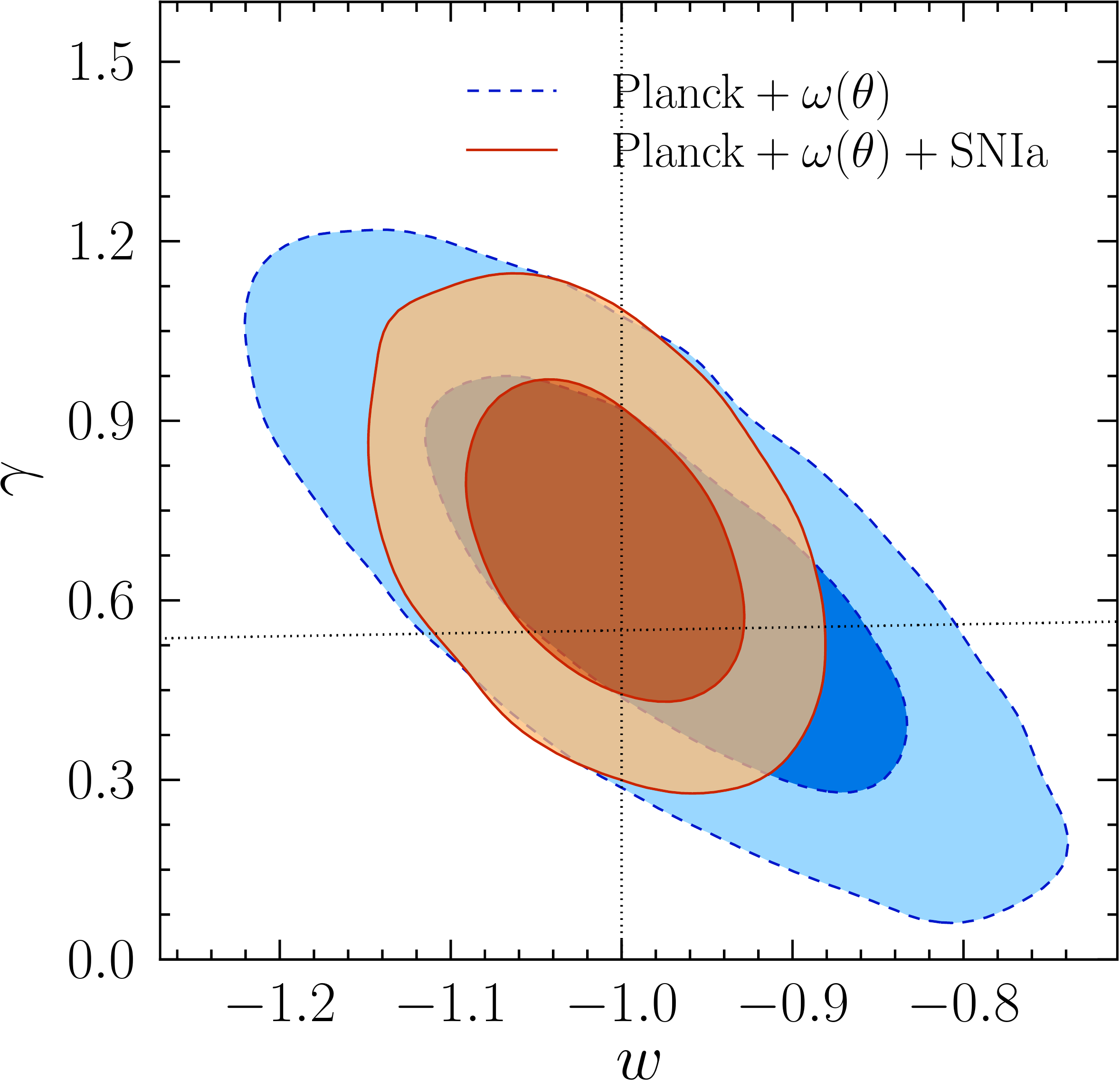}
   \caption{Marginalised $68\%$ and $95\%$ confidence interval constraints in the $w-\gamma$ plane. The blue dashed line corresponds to the constraints obtained by the $\rm{Planck}+\omega(\theta)$ combination, and the solid orange line line to those obtained combining $\rm{Planck}+\omega(\theta)+\rm{SNIa}$. The (almost) horizontal dotted line shows the value of $\gamma$ that recovers the GR prediction for the growth rate $f$, following equation (\ref{eq:gammaGR}).}
   \label{fig:wgamma}
\end{figure}

First, we assume the standard $\Lambda$CDM as the background cosmological model. A summary of the obtained constraints can be found in table \ref{tab:glcdm}. Figure \ref{fig:Omgamma} shows the marginalised 68 and 95 per cent confidence interval constraints in the $\Omega_{\rm m}-\gamma$ plane. Since CMB cannot be used to measure $f(z)$ and thus constrain $\gamma$, Planck-only contours are not shown, and the blue dashed line corresponds to the constraints obtained by the $\rm{Planck}+\omega(\theta)$ combination, while the solid orange line line to those obtained combining $\rm{Planck}+\omega(\theta)+\rm{SNIa}$. For the former we find $\Omega_{\rm m} = 0.317^{+0.011}_{-0.013}$ and $\gamma = 0.67\pm 0.15$. Then, similar to what we obtain for the $\Lambda$CDM results, adding SNIa does not significantly improve the constraints, resulting in $\Omega_{\rm m} = 0.315\pm0.011$ and $\gamma = 0.68\pm 0.14$. Both data-set combinations result in constraints that are in good agreement with GR within $1\sigma$, as well as with our previous results for the basic $\Lambda$CDM case.

Finally, constraints obtained also treating $w_{\rm{DE}}$ as a free parameter, assuming that it is constant in time, are listed in table \ref{tab:gwcdm}. Figure \ref{fig:wgamma} shows the marginalised 68 and 95 per cent confidence interval constraints in the $w-\gamma$ plane. The vertical dotted line marks $w=-1$, while the other one follows equation (\ref{eq:gammaGR}). Using the $\rm{Planck}+\omega(\theta)$ combination, we obtain a value of \break$w=-0.980\pm 0.092$ and $\gamma = 0.64^{+0.21}_{-0.23}$. Adding the information from SNIa tightens the constraints, resulting in $w=-1.013^{+0.052}_{-0.047}$ and $\gamma = 0.70^{+0.16}_{-0.18}$. Both sets of constraints are again in good agreement with the standard $\Lambda$CDM model and General Relativity.

\subsection{Comparison to companion papers}

This work forms part of a collective effort consisting of a number of different analyses of the completed BOSS combined sample, which compliment, support and converge in \cite{Alam:2016aa}. Full-shape (FS) anisotropic clustering measurements, in 3D configuration-space as well as in Fourier-space, are presented in \cite{Beutler:2017ab, Grieb:2017aa, Sanchez:2017aa, Satpathy:2016aa}. Anisotropic BAO-only measurements post-reconstruction \citep{Eisenstein:2007aa,Padmanabhan:2012aa} are presented in \cite{Beutler:2017aa, Ross:2017aa, Vargas-Magana:2017aa}. All these different methods are optimally combined following the method exposed in \cite{Sanchez:2017ab}, and used in \cite{Alam:2016aa} to constraint a variety of parameter spaces for different cosmological models.

In this analysis we do not derive intermediate measurements, between the clustering of galaxies and the final cosmological constraints, as is the case of the analyses mentioned above using a more standard approach, where $D_{\rm M}$, $H$ and $f\sigma_8$ is measured. For this reason it is difficult to directly compare to our companion papers, unless we do something similar to what is done in Section \ref{sec:tests}, although this would always be in the context of the cosmological model assumed to derive distance measurements from $\omega(\theta)$. 

A more quantitative comparison would be to compare the constraints themselves for each parameter spaces explored. Although, it should be noted that in \cite{Alam:2016aa} the CMB data used also includes high-$\ell$ $E$-mode polarisation auto-spectra, as well as high-$\ell$ $TE$ cross-spectra, while we have limited our analysis to only use the base case from CMB observations.

In general, the constraints on cosmological parameters presented in \cite{Alam:2016aa} are highly consistent with those presented in this analysis, but tighter by $\sim40\%$. This difference is expected not only from the extra CMB information, but also from the fact that the results presented there benefit from the combination of four FS plus two BAO-only measurements. We note however, that in the cases where alternative dark-energy models are explored this difference in precision is significantly reduced, confirming that this tomographic technique can provide strong constraints on the expansion history of the Universe.

Special mention should be made to two of our companion papers, \cite{Wang:2016aa} and \cite{Zhao:2017aa}, whose analyses are particularly complementary to ours. Both papers perform anisotropic BAO-only measurements in tomographic bins, both in 3D configuration-space \citep{Wang:2016aa} and Fourier-space \citep{Zhao:2017aa}. Similarities are evident, to perform tomographic clustering measurements in many redshift bins to leverage the information enclosed in the time evolution of the clustering signal. The main differences are, firstly, these two analyses use the 3D position of galaxies, which does not suffer from projection effects as the technique in our analysis, increasing the significance of the BAO detection at expenses of assuming a fiducial cosmology for their measurements, which is one of the strengths of our analysis. Secondly, \cite{Wang:2016aa} and \cite{Zhao:2017aa} perform anisotropic BAO-only fits, while in this analysis we model the full shape of $\omega(\theta)$, which encodes information of the growth of structures. In overall, this is reflected in that our analysis provides tighter constraints on $\Omega_{\rm m}$ and can constrain growth-related parameters such as the growth index $\gamma$ in Section \ref{sec:gcdm}, while the other two analyses provide tighter constraints in parameters related to the expansion history of the Universe, such as $w_0$ and $w_a$. (see e.g. Table 7 in \citealt{Wang:2016aa}).

\section{Conclusions}\label{sec:conclusions}
We applied a tomographic technique to analyse galaxy clustering based on \cite{Salazar-Albornoz:2014aa} to the final BOSS galaxy sample. For this purpose, we extended our description of the full shape of $\omega(\theta)$ to use state-of-the-art modelling of non-linearities, galaxy bias and RSD. We also extended the analysis to include cross-correlation measurements between redshift shells. 

In order to maximise the constraining power of our measurements, we optimised the number of redshift shells used in the analysis, by means of maximising the FoM in the $\Omega_{\rm m}-w$ plane. We did this exploring three different cases: (i) a Fisher-matrix approach that resulted in an monotonic increase in the FoM as a function of the number of shells; (ii) an MCMC analysis using synthetic data, where we only varied $\Omega_{\rm m}$ and $w$, which showed a clear maximum in the FoM; and (iii), an analogous MCMC test, where we also included the nuisance parameters of the model, which resulted in the same behaviour as (ii), but with a smaller value for the FoM. We defined our binning scheme on the basis of the last case, where our final configuration consisted of 18 redshift shells of different widths, containing $\sim70000$ galaxies each, plus as many cross-correlations, with subsequent shells, as necessary to surpass the BAO scale in the line of sight. 

We tested our methodology against a set of $1000$ {\sc md-patchy} mock catalogues, which are designed to match the characteristics of the final BOSS galaxy sample, following its angular and radial selection function, as well as including the redshift evolution of bias and RSD. Using the mean of the $1000$ mock catalogues, we ran an MCMC analysis constraining very general cosmologies, using three different models for the evolution of the linear galaxy-bias. We were able to recover unbiased cosmological information for two of these models, and biased results at the $1\sigma$ level for the constant galaxy-clustering (CGC) model. Also, we repeated this test on a subset of $100$ mocks using one of the galaxy-bias models that resulted in unbiased constraints, and performed an MCMC analysis on each mock catalogue individually. On these tests we found excellent agreement between the statistical errors and those estimated by our model for the full covariance matrix of $\omega(\theta)$.

Next, we analysed the redshift evolution of the linear bias of BOSS galaxies. Fixing the cosmological parameters to the best-fitting $\Lambda$CDM model to the final Planck CMB observations, we fit the linear bias parameter of our model for the galaxy-clustering signal, marginalising over the other nuisance parameters and $\sigma_8$ with a Planck prior. Also, using the same three different models for the redshift evolution of the linear galaxy-bias used in the previous tests, we fit the clustering amplitude of $\omega(\theta)$ in all redshift shells simultaneously. We saw that all three models are able to reproduce well the observed redshift evolution of the linear bias up to redshift $z\sim0.6$, where the BOSS sample is close to a volume-limited one. However, none of them were able to reproduce the observed scatter in the measurements within $0.6\lesssim z\lesssim 0.75$, where the BOSS sample behaves as flux-limited. For this reason, and because two of the three bias models depend on the linear growth factor $D(z)$, in order to avoid biased cosmological constraints, we decided not to include the measurements in these high-redshift shells in our tomographic analysis. We tested the impact that assuming these three models for the redshift evolution of the linear galaxy-bias has on the obtained constraints on cosmological parameters. Combining our measurements of $\omega(\theta)$ from BOSS with the CMB measurements from Planck, we obtained constraints on the $w$CDM parameter-space using each of the three galaxy-bias models, and found no significant difference between them, showing that this analysis provides robust constraints. 

Finally, combining the information obtained from the application of our tomographic approach to the final BOSS galaxy-sample, with the latest Planck CMB observations and type Ia supernova (SNIa), we constrain the parameters of the standard $\Lambda$CDM cosmological model and its more important extensions, including non-flat universes, more general dark-energy models, neutrino masses, and possible deviations from the predictions of general relativity. In general, these constraints are comparable to the most precise present-day cosmological constraints in the literature, showing and consolidating the $\Lambda$CDM model as the standard cosmological paradigm. 

In particular, in all the cases where we allow $w_{\rm{DE}}$ to deviate from its fiducial value of $-1$, either as constant or time-dependent, our final constraints are in good agreement to those cases where $w_{\rm{DE}}$ is fixed to $-1$. For the simplest $w$CDM extension we obtain $w_{\rm{DE}} = -0.958^{+0.063}_{-0.055}$ for the combination of our $\omega(\theta)$ measurements with Planck, and $w_{\rm{DE}} = -0.991\pm 0.046$ for the full $\rm{Planck}+\omega(\theta)+\rm{SNIa}$ combination. For models including $\Omega_{\rm K}$, with $w$ fixed to $-1$ or treated as a free parameter, we find $|\Omega_{\rm K}|\sim10^{-3}$, consistent with no curvature within the errors. Although we do not find a clear detection for the total sum of neutrino masses, we obtain upper limits that can be considered among the tightest ones available at present, where in the $\nu\Lambda$CDM case, we obtain $\sum m_\nu/\rm{eV}< 0.207(0.400)$ $68\%$($95\%$) confidence interval(C.I.) upper limits for the $\rm{Planck}+\omega(\theta)$ combination, while for the full $\rm{Planck}+\omega(\theta)+\rm{SNIa}$ case, we find $\sum m_\nu/\rm{eV}< 0.169(0.330)$ $68\%$($95\%$) C.I. upper limits. Furthermore, we see no significant deviations from the GR predictions of the linear growth of structures, parametrised by the growth index parameter $\gamma$, neither assuming a $\Lambda$CDM as the background cosmological model, nor when we also treat $w_{\rm{DE}}$ as a free parameter.

In summary, the methodology of analysing the large-scale structure of the Universe presented in this work, using angular galaxy-clustering measurements in thin redshift-shells, is an excellent alternative to the traditional 3D clustering analysis. It avoids the two main issues of the traditional approach, by using cosmology-independent measurements, and by being able to trace the redshift evolution of the clustering signal. Furthermore, this technique is able to provide precise constraints on cosmological parameters, proving to be a valid and very robust method to analyse present and future large galaxy-surveys.

\section*{Acknowledgments}
We would like to thank Daniel Farrow, Martha Lippich, Francesco Montesano and Jiamin Hou for useful discussions. SS-A, AGS and JNG acknowledge support by the Trans-regional Collaborative Research Centre TRR33 {\it The Dark Universe} of the Deutsche Forgschungsgemeinschaft (DFG). 

Funding for SDSS-III has been provided by the Alfred P. Sloan Foundation, the Participating Institutions, the National Science Foundation and the US Department of Energy.

SDSS-III is managed by the Astrophysical Research Consortium for the Participating Institutions of the SDSS-III Collaboration including the University of Arizona, the Brazilian Participation Group, Brookhaven National Laboratory, University of Cambridge, Carnegie Mellon University, University of Florida, the French Participation Group, the German Participation Group, Harvard University, the Instituto de Astrof\'isica de Canarias, the Michigan State/Notre Dame/JINA Participation Group, Johns Hopkins University, Lawrence Berkeley National Laboratory, Max Planck Institute for Astrophysics, Max Planck Institute for Extraterrestrial Physics, New Mexico State University, New York University, Ohio State University, Pennsylvania State University, University of Portsmouth, Princeton University, the Spanish Participation Group, University of Tokyo, University of Utah, Vanderbilt University, University of Virginia, University of Washington and Yale University.

This work is based on observations obtained with Planck (http://www.esa.int/Planck), an ESA science mission with instruments and contributions directly funded by ESA Member States, NASA and Canada.

Some of the results in this work have been derived using the {\sc HEALPix} \cite{Gorski:2005aa} package.

\appendix
\section{Summary tables}\label{sec:tables}

This section contains the summary tables displaying the final binning scheme used in this study (see Section \ref{sec:opti}), as well as the cosmological parameters explored in Section  \ref{sec:constraints}, that have been removed from the body of this paper for readability.

\begin{table}
\caption{Marginalised constraints on the cosmological parameters for the $\Lambda$CDM model. Values correspond to the mean and $68\%$ confidence interval. The first block corresponds to varied parameters in the analysis, while the second block are derived parameters.}
\begin{center}
\begin{tabular} {| l | c | c |}
\hline
 Parameter &  CMB + $\omega(\theta)$ & CMB + $\omega(\theta)$ + SNIa \\
\hline\hline
{ $\Omega_b h^2   $} 		& $0.02215\pm 0.00021$  & $0.02217\pm 0.00021$\\
                                                               
{ $\Omega_c h^2   $} 		& $0.1204\pm 0.0019  $  	& $0.1200\pm 0.0018  $\\
                                                               
{ $100\theta_{\rm{MC}} $} 		& $1.04078\pm 0.00045$  & $1.04080\pm 0.00043$\\
                                                               
{ $\tau           $} 			& $0.070\pm 0.018    $  	& $0.072\pm 0.018    $\\
                                                               
{ ${\rm{ln}}(10^{10} A_s)$} 	& $3.075\pm 0.034    $  	& $3.077\pm 0.035    $\\
                                                               
{ $n_s            $} 			& $0.9631\pm 0.0053  $  	& $0.9637\pm 0.0053  $\\

\hline                                                               

$H_0                       $ 			& $66.98\pm 0.80     $  	& $67.14\pm 0.77     $\\
                                                               
$\Omega_\Lambda            $ 		& $0.681\pm 0.011    $  	& $0.683\pm 0.011    $\\
                                                               
$\Omega_{\rm m}                  $ 			& $0.319\pm 0.011    $  	& $0.317\pm 0.011    $\\
                                                               
$\sigma_8                  $ 			& $0.825\pm 0.014    $  	& $0.825\pm 0.014    $\\
                                                               
${\rm{Age}}/{\rm{Gyr}}     $ 		& $13.826\pm 0.033   $  	& $13.822\pm 0.032   $\\
\hline
\end{tabular} 
\label{tab:lcdm}
\end{center}
\end{table}

 \begin{table}
\caption{Marginalised constraints on the cosmological parameters for the $w$CDM model. Values correspond to the mean and $68\%$ confidence interval. The first block corresponds to varied parameters in the analysis, while the second block are derived parameters.}
\begin{center}
\begin{tabular} {| l | c | c |}
\hline
 Parameter &  CMB + $\omega(\theta)$ & CMB + $\omega(\theta)$ + SNIa \\
\hline\hline
{ $\Omega_b h^2   $} 		& $0.02220\pm 0.00022      $ 	& $0.02219\pm 0.00022     $\\
                                                                     
{ $\Omega_c h^2   $} 		& $0.1198\pm 0.0021        $ 	& $0.1199\pm 0.0021       $\\
                                                                     
{ $100\theta_{\rm{MC}} $} 		& $1.04087\pm 0.00045      $ 	& $1.04085\pm 0.00046     $\\
                                                                     
{ $\tau           $} 			& $0.076\pm 0.019          $ 	& $0.074\pm 0.019         $\\
                                                                     
{ $w              $} 			& $-0.958^{+0.063}_{-0.055}$ 	& $-0.991\pm 0.046        $\\
                                                                     
{ ${\rm{ln}}(10^{10} A_s)$} 	& $3.087\pm 0.037          $ 	& $3.081\pm 0.036         $\\
                                                                     
{ $n_s            $} 			& $0.9647\pm 0.0059        $ 	& $0.9645\pm 0.0059       $\\

\hline
                                                                     
$H_0                       $ 			& $66.0\pm 1.5             $ 		& $66.9\pm 1.1            $\\
                                                                     
$\Omega_\Lambda            $ 		& $0.672\pm 0.016          $ 	& $0.681^{+0.013}_{-0.011}$\\
                                                                     
$\Omega_{\rm m}                  $ 			& $0.328\pm 0.016          $ 	& $0.319\pm 0.012         $\\
                                                                     
$\sigma_8                  $ 			& $0.816\pm 0.020          $ 	& $0.823\pm 0.019         $\\
                                                                     
${\rm{Age}}/{\rm{Gyr}}     $ 		& $13.844\pm 0.040         $ 	& $13.825\pm 0.034        $\\
\hline
\end{tabular}
\label{tab:wcdm}
\end{center}
\end{table}

 \begin{table}
\caption{Marginalised constraints on the cosmological parameters for the $w_0w_a$CDM model. Values correspond to the mean and $68\%$ confidence interval. The first block corresponds to varied parameters in the analysis, while the second block are derived parameters.}
\begin{center}
\begin{tabular} {| l | c | c |}
\hline
 Parameter &  CMB + $\omega(\theta)$ & CMB + $\omega(\theta)$ + SNIa\\
\hline\hline
{ $\Omega_b h^2   $} 		& $0.02220\pm 0.00022     $  & $0.02216\pm 0.00022        $\\

{ $\Omega_c h^2   $} 		& $0.1199\pm 0.0022       $  & $0.1199\pm 0.0021          $\\

{ $100\theta_{\rm{MC}} $} 		& $1.04084\pm 0.00048     $  & $1.04084\pm 0.00044        $\\

{ $\tau           $} 		& $0.076\pm 0.019         $  & $0.074\pm 0.019            $\\

{ $w_0              $} 		& $-0.60^{+0.24}_{-0.10}  $  & $-0.94\pm 0.13             $\\

{ $w_a            $} 		& $< -0.965               $  & $-0.23^{+0.51}_{-0.42}     $\\

{ ${\rm{ln}}(10^{10} A_s)$}	& $3.087\pm 0.036         $  & $3.082\pm 0.036            $\\

{ $n_s            $} 		& $0.9647\pm 0.0061       $  & $0.9637\pm 0.0060          $\\

\hline

$H_0                       $ 		& $64.3^{+1.3}_{-1.8}     $  & $67.0\pm 1.2               $\\

$\Omega_\Lambda            $ 		& $0.654^{+0.017}_{-0.019}$  & $0.681\pm 0.012            $\\

$\Omega_{\rm m}                  $ 		& $0.346^{+0.019}_{-0.017}$  & $0.319\pm 0.012            $\\

$\sigma_8                  $ 		& $0.806\pm 0.021         $  & $0.825\pm 0.018            $\\

${\rm{Age}}/{\rm{Gyr}}     $ 		& $13.790\pm 0.046        $  & $13.811^{+0.047}_{-0.055}  $\\
\hline
\end{tabular}
\label{tab:wacdm}
\end{center}
\end{table}

 \begin{table}
\caption{Marginalised constraints on the cosmological parameters for the $k$CDM model. Values correspond to the mean and $68\%$ confidence interval. The first block corresponds to varied parameters in the analysis, while the second block are derived parameters.}
\begin{center}
\begin{tabular} {| l | c | c |}
\hline
  Parameter &  CMB + $\omega(\theta)$ & CMB + $\omega(\theta)$ + SNIa\\
\hline\hline
{ $\Omega_b h^2   $} 		& $0.02230\pm 0.00026         $   	& $0.02229\pm 0.00026        $\\

{ $\Omega_c h^2   $} 		& $0.1189\pm 0.0022           $   		& $0.1192^{+0.0022}_{-0.0026}$\\

{ $100\theta_{\rm{MC}} $} 		& $1.04100\pm 0.00049         $   	& $1.04102^{+0.00055}_{-0.00049}$\\

{ $\tau           $} 			& $0.076\pm 0.020             $   		& $0.074^{+0.016}_{-0.021}   $\\
                                                                     
{ $\Omega_{\rm K}       $} 		& $-0.0043^{+0.0042}_{-0.0035}$   	& $-0.0028\pm 0.0038         $\\

{ ${\rm{ln}}(10^{10} A_s)$} 	& $3.085\pm 0.039             $   		& $3.080^{+0.032}_{-0.038}   $\\

{ $n_s            $} 			& $0.9671^{+0.0059}_{-0.0073} $   	& $0.9663^{+0.0071}_{-0.0061}$\\

\hline

$H_0                       $ 		& $65.7^{+1.5}_{-1.3}         $   & $66.3\pm 1.2               $\\

$\Omega_\Lambda            $ 		& $0.676\pm 0.013             $   & $0.679^{+0.013}_{-0.0093}  $\\

$\Omega_{\rm m}                  $ 		& $0.329^{+0.014}_{-0.016}    $   & $0.324^{+0.011}_{-0.014}   $\\

$\sigma_8                  $ 		& $0.823\pm 0.015             $   & $0.822\pm 0.014            $\\

${\rm{Age}}/{\rm{Gyr}}     $ 		& $13.99^{+0.14}_{-0.17}      $   & $13.93\pm 0.14             $\\
\hline
\end{tabular}
\label{tab:kcdm}
\end{center}
\end{table}

 \begin{table}
\caption{Marginalised constraints on the cosmological parameters for the $wk$CDM model. Values correspond to the mean and $68\%$ confidence interval. The first block corresponds to varied parameters in the analysis, while the second block are derived parameters.}
\begin{center}
\begin{tabular} {| l | c | c |}
\hline
 Parameter &  CMB + $\omega(\theta)$ & CMB + $\omega(\theta)$ + SNIa\\
\hline\hline
{ $\Omega_b h^2   $} 		& $0.02227\pm 0.00025         $   & $0.02230\pm 0.00024         $\\
                                                                                                        
{ $\Omega_c h^2   $} 		& $0.1193\pm 0.0022           $   & $0.1187\pm 0.0022           $\\
                                                                                                        
{ $100\theta_{\rm{MC}} $} 		& $1.04095\pm 0.00045         $   & $1.04097\pm 0.00049         $\\
                                                                                                        
{ $\tau           $} 		& $0.076\pm 0.019             $   & $0.073\pm 0.019             $\\

{ $\Omega_{\rm K}       $} 		& $-0.0037^{+0.0057}_{-0.0051}$   & $-0.0040^{+0.0054}_{-0.0041}$\\

{ $w              $} 		& $-1.00^{+0.10}_{-0.075}     $   & $-1.025^{+0.064}_{-0.055}   $\\
                                                                                                        
{ ${\rm{ln}}(10^{10} A_s)$} 	& $3.084\pm 0.037             $   & $3.077\pm 0.036             $\\
                                                                                                        
{ $n_s            $} 		& $0.9657\pm 0.0064           $   & $0.9675\pm 0.0063           $\\

\hline
                                                                                                        
$H_0                       $ 		& $65.7^{+1.3}_{-1.5}         $   & $66.5\pm 1.3                $\\
                                                                                                        
$\Omega_\Lambda            $ 		& $0.673\pm 0.017             $   & $0.684\pm 0.013             $\\
                                                                                                        
$\Omega_{\rm m}                  $ 		& $0.330\pm 0.016             $   & $0.320\pm 0.014             $\\
                                                                                                        
$\sigma_8                  $ 		& $0.822\pm 0.023             $   & $0.825\pm 0.019             $\\
                                                                                                        
${\rm{Age}}/{\rm{Gyr}}     $ 		& $13.99^{+0.17}_{-0.22}      $   & $13.98^{+0.16}_{-0.21}      $\\
\hline                                                                                                  
\end{tabular}
\label{tab:wkcdm}
\end{center}
\end{table}

 \begin{table}
\caption{Marginalised constraints on the cosmological parameters for the $\nu\Lambda$CDM model. Values correspond to the mean and $68\%$ confidence interval (C.I.), except for the sum of neutrino masses where $95\%$ C.I. upper limits are shown (for $68\%$ C.I. see text). The first block corresponds to varied parameters in the analysis, while the second block are derived parameters.}
\begin{center}
\begin{tabular} {| l | c | c |}
\hline
 Parameter &  CMB + $\omega(\theta)$ & CMB + $\omega(\theta)$ + SNIa\\
\hline\hline
{ $\Omega_b h^2   $} 		& $0.02214\pm 0.00021      $  & $0.02219\pm 0.00021      $\\

{ $\Omega_c h^2   $} 		& $0.1200\pm 0.0020        $  & $0.1197\pm 0.0019        $\\

{ $100\theta_{\rm{MC}} $} 		& $1.04079\pm 0.00044      $  & $1.04085\pm 0.00045      $\\

{ $\tau           $} 		& $0.076\pm 0.019          $  & $0.077\pm 0.019          $\\

{ $\Sigma m_\nu /\rm{eV}$} 		& $< 0.400(95\%\mbox{C.I.}) $  & $< 0.330(95\%\mbox{C.I.}) $\\

{ ${\rm{ln}}(10^{10} A_s)$} 	& $3.086\pm 0.037          $  & $3.087\pm 0.037          $\\

{ $n_s            $} 		& $0.9633\pm 0.0055        $  & $0.9643\pm 0.0054        $\\

\hline

$H_0                       $ 		& $66.2^{+1.2}_{-1.0}      $  & $66.6^{+1.1}_{-0.93}     $\\

$\Omega_\Lambda            $ 		& $0.671^{+0.017}_{-0.013} $  & $0.677^{+0.015}_{-0.012} $\\

$\Omega_{\rm m}                  $ 		& $0.329^{+0.013}_{-0.017} $  & $0.323^{+0.012}_{-0.015} $\\

$\sigma_8                  $ 		& $0.804^{+0.031}_{-0.023} $  & $0.810^{+0.028}_{-0.019} $\\

${\rm{Age}}/{\rm{Gyr}}     $ 		& $13.876^{+0.051}_{-0.071}$  & $13.854^{+0.047}_{-0.062}$\\
\hline
\end{tabular}
\label{tab:mnulcdm}
\end{center}
\end{table}

 \begin{table}
\caption{Marginalised constraints on the cosmological parameters for the $\nu w$CDM model. Values correspond to the mean and $68\%$ confidence interval (C.I.), except for the sum of neutrino masses where $95\%$ C.I. upper limits are shown (for $68\%$ C.I. see text). The first block corresponds to varied parameters in the analysis, while the second block are derived parameters.}
\begin{center}
\begin{tabular} {| l | c | c |}
\hline
 Parameter &  CMB + $\omega(\theta)$ & CMB + $\omega(\theta)$ + SNIa\\
\hline\hline
{ $\Omega_b h^2   $} 		& $0.02216\pm 0.00023      $  & $0.02215\pm 0.00022      $\\

{ $\Omega_c h^2   $} 		& $0.1199\pm 0.0021        $  & $0.1198\pm 0.0021        $\\

{ $100\theta_{\rm{MC}} $} 		& $1.04081\pm 0.00048      $  & $1.04080\pm 0.00047      $\\

{ $\tau           $} 		& $0.078\pm 0.019          $  & $0.077\pm 0.019          $\\

{ $\Sigma m_\nu/\rm{eV}   $} 		& $< 0.486 (95\%\mbox{C.I.})$  & $< 0.474 (95\%\mbox{C.I.})$\\

{ $w              $} 		& $-0.998^{+0.097}_{-0.064}$  & $-1.023^{+0.063}_{-0.053}$\\

{ ${\rm{ln}}(10^{10} A_s)$} 	& $3.090\pm 0.037          $  & $3.086\pm 0.036          $\\

{ $n_s            $} 		& $0.9636\pm 0.0064        $  & $0.9635\pm 0.0060        $\\

\hline

$H_0                       $ 		& $66.1^{+1.5}_{-1.7}      $  & $66.7\pm 1.1             $\\

$\Omega_\Lambda            $ 		& $0.670\pm 0.017          $  & $0.676^{+0.015}_{-0.013} $\\

$\Omega_{\rm m}                  $ 		& $0.330\pm 0.017          $  & $0.324^{+0.013}_{-0.015} $\\

$\sigma_8                  $ 		& $0.801^{+0.028}_{-0.024} $  & $0.805^{+0.030}_{-0.024} $\\

${\rm{Age}}/{\rm{Gyr}}     $ 		& $13.882^{+0.054}_{-0.067}$  & $13.871^{+0.051}_{-0.072}$\\
\hline
\end{tabular}
\label{tab:mnuwcdm}
\end{center}
\end{table}

 \begin{table}
\caption{Marginalised constraints on the cosmological parameters for the $\gamma\Lambda$CDM model. Values correspond to the mean and $68\%$ confidence interval. The first block corresponds to varied parameters in the analysis, while the second block are derived parameters.}
\begin{center}
\begin{tabular} {| l | c | c |}
\hline
 Parameter &  CMB + $\omega(\theta)$ & CMB + $\omega(\theta)$ + SNIa\\
\hline\hline
{ $\Omega_b h^2   $} 		& $0.02219\pm 0.00022     $   & $0.02221\pm 0.00021$\\

{ $\Omega_c h^2   $} 		& $0.1201\pm 0.0020       $   & $0.1197\pm 0.0019  $\\

{ $100\theta_{\rm{MC}} $} 		& $1.04084\pm 0.00046     $   & $1.04088\pm 0.00045$\\

{ $\tau           $} 		& $0.075\pm 0.019         $   & $0.077\pm 0.019    $\\

{ ${\rm{ln}}(10^{10} A_s)$} 	& $3.084\pm 0.036         $   & $3.086\pm 0.036    $\\

{ $n_s            $} 		& $0.9641\pm 0.0057       $   & $0.9650\pm 0.0055  $\\

{ $\gamma         $} 		& $0.67\pm 0.15           $   & $0.68\pm 0.14      $\\

\hline

$H_0                       $ 		& $67.15\pm 0.87          $   & $67.33\pm 0.82     $\\

$\Omega_\Lambda            $ 		& $0.683^{+0.013}_{-0.011}$   & $0.685\pm 0.011    $\\

$\Omega_{\rm m}                  $ 		& $0.317^{+0.011}_{-0.013}$   & $0.315\pm 0.011    $\\

$\sigma_8                  $ 		& $0.828\pm 0.014         $   & $0.828\pm 0.015    $\\

${\rm{Age}}/{\rm{Gyr}}     $ 		& $13.819\pm 0.036        $   & $13.813\pm 0.034   $\\
\hline
\end{tabular}
\label{tab:glcdm}
\end{center}
\end{table}

 \begin{table}
\caption{Marginalised constraints on the cosmological parameters for the $\gamma w$CDM model. Values correspond to the mean and $68\%$ confidence interval. The first block corresponds to varied parameters in the analysis, while the second block are derived parameters.}
\begin{center}
\begin{tabular} {| l | c | c |}
\hline
 Parameter &  CMB + $\omega(\theta)$ & CMB + $\omega(\theta)$ + SNIa\\
\hline\hline
{ $\Omega_b h^2   $} 		& $0.02220\pm 0.00022      $  & $0.02220\pm 0.00022      $\\

{ $\Omega_c h^2   $} 		& $0.1199\pm 0.0021        $  & $0.1200\pm 0.0020        $\\

{ $100\theta_{\rm{MC}} $} 		& $1.04086\pm 0.00046      $  & $1.04088\pm 0.00045      $\\

{ $\tau           $} 		& $0.076\pm 0.019          $  & $0.076\pm 0.019          $\\

{ $w              $} 		& $-0.980\pm 0.092         $  & $-1.013^{+0.052}_{-0.047}$\\

{ ${\rm{ln}}(10^{10} A_s)$} 	& $3.086\pm 0.037          $  & $3.086\pm 0.036          $\\

{ $n_s            $} 		& $0.9644\pm 0.0060        $  & $0.9643\pm 0.0059        $\\

{ $\gamma         $} 		& $0.64^{+0.21}_{-0.23}    $  & $0.70^{+0.16}_{-0.18}    $\\

\hline

$H_0                       $ 		& $66.6\pm 2.5             $  & $67.6\pm 1.3             $\\

$\Omega_\Lambda            $ 		& $0.677^{+0.027}_{-0.022} $  & $0.687\pm 0.013          $\\

$\Omega_{\rm m}                  $ 		& $0.323^{+0.022}_{-0.027} $  & $0.313\pm 0.013          $\\

$\sigma_8                  $ 		& $0.822\pm 0.030          $  & $0.832\pm 0.020          $\\

${\rm{Age}}/{\rm{Gyr}}     $ 		& $13.832^{+0.053}_{-0.064}$  & $13.809\pm 0.037         $\\
\hline
\end{tabular}
\label{tab:gwcdm}
\end{center}
\end{table}

\begin{table}
\caption{Redshift limits and $\Delta z$ of the $18$ z-shells found to be the optimal binning scheme for this tomographic analysis of this paper, form which the three higher redshift were not used. In all the Figures, the redshift limits are shown only to three decimal points.}
\begin{center}
\begin{tabular}{|c|c|c|c|}
\hline
$z_\rmn{min}$ & $z_\rmn{max}$ & $\Delta z$ & \mbox{used}\\
\hline\hline
$0.20000$ & $0.25841$ &  $0.05841$ & \mbox{yes}\\
$0.25841$ & $0.30813$ &  $0.04972$ & \mbox{yes}\\
$0.30813$ & $0.34266$ &  $0.03453$ & \mbox{yes}\\
$0.34266$ & $0.37622$ &  $0.03356$ & \mbox{yes}\\
$0.37622$ & $0.41421$ &  $0.03799$ & \mbox{yes}\\
$0.41421$ & $0.44550$ &  $0.03129$ & \mbox{yes}\\
$0.44550$ & $0.46670$ &  $0.02121$ & \mbox{yes}\\
$0.46670$ & $0.48305$ &  $0.01635$ & \mbox{yes}\\
$0.48305$ & $0.49783$ &  $0.01478$ & \mbox{yes}\\
$0.49783$ & $0.51177$ &  $0.01394$ & \mbox{yes}\\
$0.51177$ & $0.52580$ &  $0.01403$ & \mbox{yes}\\
$0.52580$ & $0.54021$ &  $0.01442$ & \mbox{yes}\\
$0.54021$ & $0.55550$ &  $0.01529$ & \mbox{yes}\\
$0.55550$ & $0.57185$ &  $0.01635$ & \mbox{yes}\\
$0.57185$ & $0.59103$ &  $0.01918$ & \mbox{yes}\\
\hline
$0.59103$ & $0.61356$ &  $0.02253$ & \mbox{no}\\
$0.61356$ & $0.64375$ &  $0.03018$ & \mbox{no}\\
$0.64375$ & $0.75000$ &  $0.10625$ & \mbox{no}\\
\hline
\end{tabular}
\label{tab:zlim}
\end{center}
\end{table}
\onecolumn

\begin{figure*}
   \includegraphics[scale=0.75]{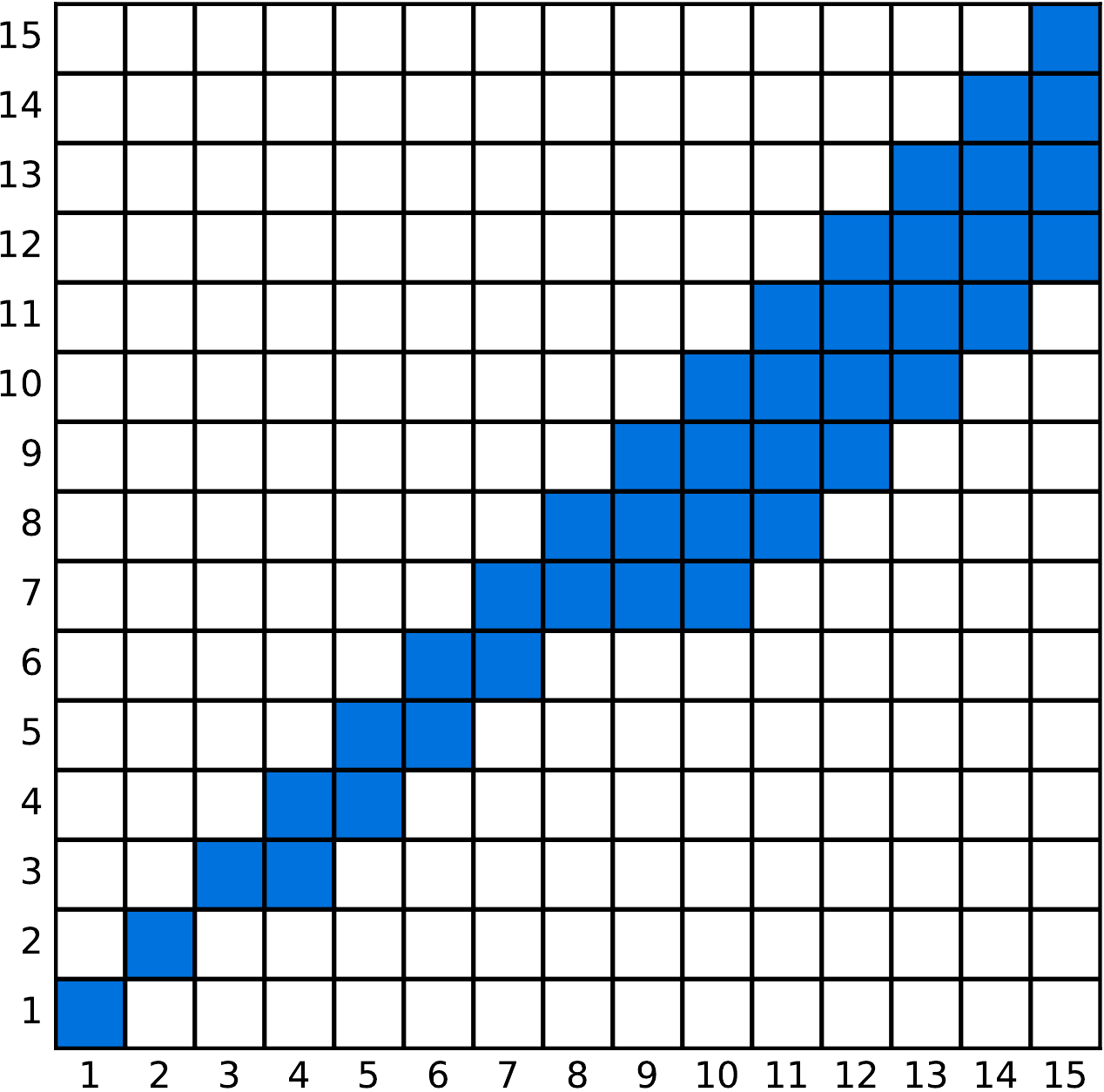}
   \caption{Configuration matrix illustrating the auto- and cross-correlation functions used in the analysis of the cosmological implications of $\omega(\theta)$ measured on BOSS. Filled entries indicate the measurements used, where the diagonal terms are the auto-correlations, and off-diagonal terms correspond to the cross-correlations included.}
   \label{fig:confmat}
\end{figure*}

\twocolumn
\bibliographystyle{mnras}
\bibliography{draft}

\label{lastpage}

\end{document}